\documentclass[apl,amsmath,amssymb,reprint,longbibliography,showkeys,superscriptaddress,balancelastpage]{revtex4-1}

\usepackage{textcomp}
\usepackage{graphicx}
\usepackage{dcolumn}
\usepackage{bm}
\usepackage{multirow}
\usepackage{gensymb} 
\usepackage[export]{adjustbox}
\usepackage{pifont}

\makeatletter
\def\NAT@def@citea{\def\@citea{\NAT@separator}}
\makeatother

\usepackage{color}
\usepackage{cleveref}
\crefformat{figure}{#2Fig.~#1#3}
\crefformat{equation}{#2Eq.~(#1)#3}
\crefformat{table}{#2Table~#1#3}
\crefformat{section}{#2Section~#1#3}

\usepackage{verbatim}

\begin{document}

\title{Towards quantitative inference of nanoscale defects in irradiated metals and alloys}

\author{Charles A. Hirst}
	\email{cahirst@mit.edu}
	\affiliation{Department of Nuclear Science and Engineering, Massachusetts Institute of Technology, Cambridge, MA, USA}
\author{Cody A. Dennett}
	\email{cdennett@mit.edu}
	\affiliation{Department of Nuclear Science and Engineering, Massachusetts Institute of Technology, Cambridge, MA, USA}
	\affiliation{Materials Science and Engineering Department, Idaho National Laboratory, Idaho Falls, ID, USA}

\date{\today}

\begin{abstract}

Quantifying the population of nanoscale defects that are formed in metals and alloys exposed to extreme radiation environments remains a pressing challenge in materials science. These defects both fundamentally alter material properties and seed long-timescale performance degradation, which often limits the lifespan of engineering systems. Unlike ceramic and semiconducting materials, these defects in metals and alloys are not spectroscopically active, forcing characterization to rely on indirect measurements from which the distribution of nanoscale defects may be inferred. In this mini-review, different experimental methodologies which have been employed for defect inference are highlighted to capture the current state of the art. Future directions in this area are proposed, which, by combining data streams from multiple and complementary characterization methods in concert with multi-scale modeling and simulation, will enable the ultimate goal of quantifying the full spectrum of defects in irradiated metals and alloys.

\end{abstract}


\maketitle

\section{Introduction} 
Materials in extreme radiation environments -- from nuclear energy systems, to particle accelerators, to satellites -- experience some of the most demanding sets of conditions for components in-service~\cite{Allen2010,Gilbert2021}. In addition to elevated temperatures, stresses, and corrosive species, materials must withstand fluxes of high energy particles, often over long operational lifetimes. These particles collide with atoms, creating cascades of displacements on very short timescales~\cite{Zinkle1993}. The formation and evolution of these primary, nanoscale defects eventually lead to microstructural changes across all length scales and results in the degradation of material properties.

Techniques used to investigate structural defects in irradiated materials include, but are not limited to, electron microscopy~\cite{Jenkins2001}, optical spectroscopies~\cite{Rickert2022}, X-ray and neutron scattering~\cite{Ehrhart1994,Albertini1992}, ion-beam analysis~\cite{Swanson1982}, field ion microscopy~\cite{Seidman1978}, and positron annihilation spectroscopy~\cite{Selim2021}. While transmission electron microscopy (TEM) has been extensively used to characterize radiation-affected microstructures~\cite{Jenkins1994}, it is fundamentally unable to detect the full spectrum of defects in a material~\cite{Jenkins2001} due to a practical resolution limit of $\sim$1~nm~\cite{Zhou2007}. Simulations show that displacement cascades create a power law-scaled distribution of defect clusters~\cite{Yi2015}, implying that the smallest defects are the most prevalent in irradiated metals. A void with a diameter of 1~nm corresponds to a cluster of $\sim$350 vacancies~\cite{Caturla2000}. With TEM unable to reliably resolve clusters of this size and below, electron microscopy drastically underestimates the total defect density in irradiated materials~\cite{Meslin2010,Reza2020,Ungar2021}.

In order to understand the mechanisms which govern irradiation-induced changes in properties, it is crucial to characterize the formation and evolution of defects on these smallest scales. In addition to seeding larger scale defect formation, nanoscale defects can have a significant effect on the properties of irradiated materials. Reza et al. show that defects below the resolution limit of TEM play a dominant role in the decrease of thermal diffusivity for self-ion irradiated W~\cite{Reza2020}. Li et al. report that the ultrahigh hardening of He-irradiated Nb results from the presence of vacancies and He-V complexes~\cite{Li2022}. Thus, for accurate prediction of irradiated materials' properties at the macroscale, it is critical to characterize defects at the nanoscale.



In non-metals, these defects can often be characterized through optical spectroscopic techniques such as Raman spectroscopy~\cite{Shelyug2018}, optical ellipsometry~\cite{Khanolkar2022}, photoluminescence~\cite{Khanolkar2022}, and others. In Raman-active materials, the vibrational modes of different bonds give unique signatures that can be probed optically and provide insight into defects as small as isolated vacancies and interstitials. Similarly, these point defects in many ceramics are charged, leaving them optically accessible through absorption, excitation, or emission. Due to charge screening from free electrons in metals and alloys, these spectroscopic methods may not be applied readily to the study of defects. Thus, nanoscale defects must be detected through alternative means. Their presence, type, and density can be determined indirectly through their effect on certain material properties and structural features. The purpose of this mini-review is to highlight developments in these inference-based techniques and report progress towards the quantitative characterization of nanoscale defects in irradiated metals and alloys.

\begin{figure*}[!ht]
\includegraphics[width=0.95\textwidth]{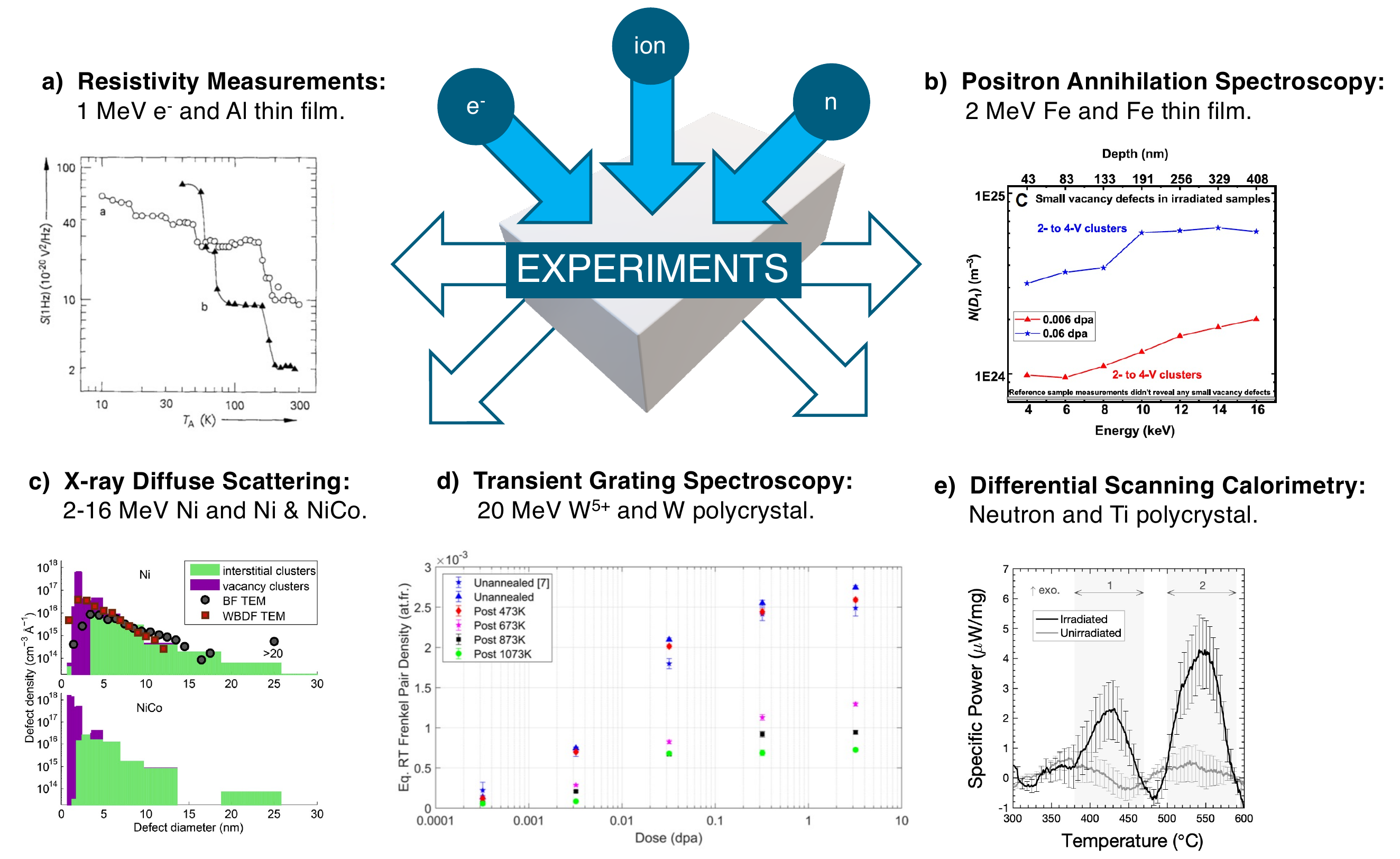}
\caption{\textbf{Many experimental techniques can be used to infer nanoscale defects in irradiated metals and alloys.} \textbf{a)} Novel resistivity measurements reveal recovery stages unseen in conventional experiments. From~\cite{Briggmann1994}, Copyright 2006 John Wiley and Sons. \textbf{b)} Variable energy PAS can resolve vacancy-type defects with depth. From~\cite{Agarwal2020}, CC BY-NC 4.0. \textbf{c)} XDS can determine the size distribution of dislocation loops down to $<$1nm. From~\cite{Olsen2016}, Copyright 2016 Elsevier. \textbf{d)} TGS coupled to kinetic theory allows estimation of the Frenkel pair defect density as a function of dose. From~\cite{Reza2021}, CC BY 4.0. \textbf{e)} DSC can characterize previously unreported defect annealing stages. From~\cite{Hirst2021}, CC BY 4.0. \label{fig:Exp}}
\end{figure*}

\section{Resistivity measurements} 

Some of the oldest methods deployed to infer populations of nanoscale defects in metals are those relying on changes in electrical resistivity. Perturbations of the crystalline lattice increase resistivity through enhanced scattering of electrons~\cite{Broom1954}. Net changes in resistivity can be expressed as the sum of contributions from all types of defects under the principle developed by Matthiessen~\cite{Matthiessen1864}, which considers contributions from each defect type through their concentration and specific resistivity.

Many studies have used resistivity measurements to infer the Frenkel pair concentration after cryogenic irradiation. These include experiments seeking to determine threshold displacement energies~\cite{Lucasson1962}, defect production rates, spontaneous relaxation volumes, and the saturation concentration of defects~\cite{Nakagawa1977,Nakagawa1979,Nakagawa1982}. With defect migration deactivated at low temperature, this method has enabled the evaluation of primary damage formation from different irradiation particles in FCC metals~\cite{Iwase1992} and Fe~\cite{Chimi2000}, and the validation of displacement cross sections in Cu~\cite{Iwamoto2015}.

Often these studies also use isochronal annealing to evaluate the recovery of specific defect populations with increasing temperature~\cite{Horak1975,Nakagawa1977,Nakagawa1982,Iwase1992,Lucasson1962,Chimi2000,Iwamoto2015}. These experiments can be coupled to kinetic Monte Carlo (kMC) simulations~\cite{Fu2004,Fluss2004} of the defect evolution to gain insight into the precise recovery mechanism, shown in figure \cref{fig:Sim}(a). Additionally, several studies have correlated resistivity to stored energy measurements in order to validate the change in defect concentration~\cite{Kinchin1958,Isebeck1966,Delaplace1968,Nicoud1968,Losehand1969}. Resistivity measurements have also been used to evaluate the effect of solutes on the recovery of radiation damage, including studies on C-doped Fe~\cite{Takaki1983} and Fe-Cr~\cite{GomezFerrer2016}.

Alternative measurement schemes have also been developed. For example, Briggmann et al. use the $1/f$ nature of noise from a resistivity measurement to characterize defects through their migration rather than through their annihilation~\cite{Briggmann1994}. This method reveals the presence of recovery stages, shown in \cref{fig:Exp}(a), that do not appear in conventional resistivity measurements, which are attributed to de-trapping of crowdions. Nikolaev has also championed the use of differential resistivity recovery measurements to decouple the effects of changing defect concentration and specific resistivity~\cite{Nikolaev2007}. This has been used to determine the effect of short range order in Fe-based alloys~\cite{Nikolaev2009}, and to deconvolve resistivity contributions from vacancies and interstitials~\cite{Nikolaev2018}.

One major limitation of this family of techniques includes the inability to directly simulate the resistivity of large defects, as density-functional theory (DFT) is computationally-limited to small supercell volumes. Additionally, many instances of deviation from Matthiessen's rule have been reported~\cite{Bass1972,Fluss2004}. The effect of defect clustering on the specific resistivity has been studied by Zinkle et al.~\cite{Zinkle1988} who report that the Frenkel pair resistivity for defects in small dislocation loops is similar to that of isolated Frenkel pairs, although this may not hold for larger loop sizes. These factors hinder the accurate determination of defect densities.

\section{Positron annihilation spectroscopy} 

Positrons annihilation spectroscopy (PAS) is able to detect open-volume in a crystalline lattice by virtue of locally reduced electron density, allowing incident positrons to probe vacancy-type defects~\cite{Selim2021}. Similar to resistivity measurements, positron annihilation lifetime spectroscopy (PALS) has been used for many years to investigate the formation and evolution of primary radiation damage in metals after irradiation at cryogenic temperatures and subsequent isochronal annealing~\cite{Mantl1978,Eldrup1997,Eldrup2003}. Decomposition of the positron lifetime spectrum into multiple components can be used to characterize the size distribution of vacancy clusters~\cite{Eldrup2003,Hu2016}, although there is a limit to the number of components that can be identified.

Additionally, the local chemical environment around vacancy-type defects can be explored through analysis of the momentum of the annihilating electron in Coincidence Doppler Broadening (CDB) experiments~\cite{Selim2021}. This method has enabled the detection of solute-vacancy complexes in electron-irradiated reactor pressure vessel (RPV) model alloys. Nagai et al.~\cite{Nagai2003} used CDB experiments to estimate the local concentration of solute around vacancies and attribute the irradiation-induced hardening of RPV steels to the formation of these features. 

By varying the energy of the incident positrons, the depth-dependence of defect populations can be resolved in materials with heterogeneous microstructures~\cite{Lynn1986,Siemek2021}. Recently, Agarwal et al.~\cite{Agarwal2020} used this method to probe the depth dependence of vacancy clusters created following 2~MeV self-ion irradiation of Fe thin-films, shown in figure \cref{fig:Exp}(b). Their work reports an increase in the density of small vacancy clusters with depth and an associated decrease in large vacancy clusters. Most strikingly, they report a decrease in the diameter of cavities with increasing dose, which is attributed to a new mechanism of interstitial-induced shrinkage of voids and resultant intra-cascade nucleation of small vacancy clusters.

\begin{figure*}[!ht]
\includegraphics[width=0.95\textwidth]{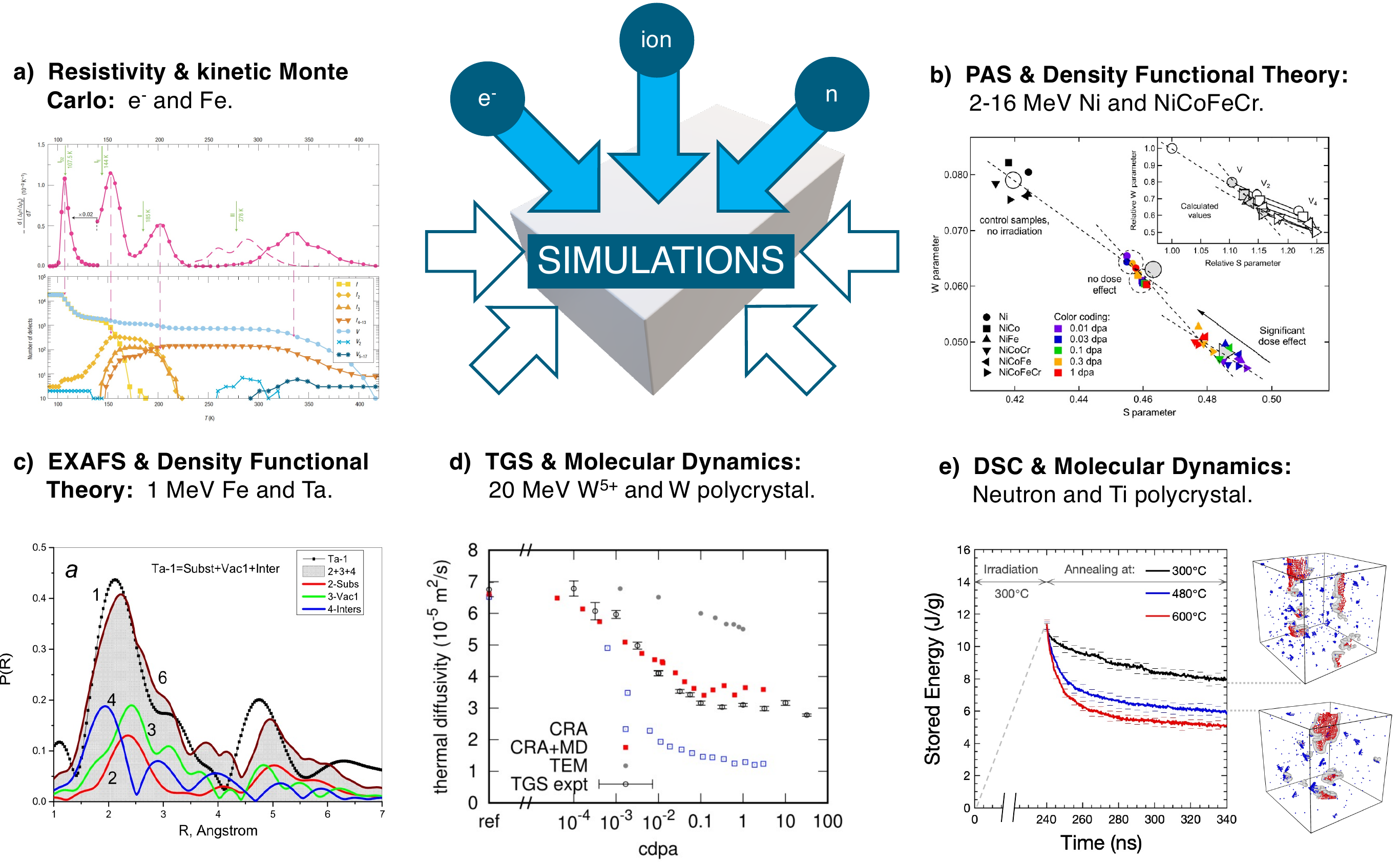}
\caption{\textbf{Simulation techniques can be used to aid the inference of nanoscale defects in irradiation experiments.} \textbf{a)} kMC can predict the isochronal evolution of defects and thus simulate the resistivity recovery spectrum. From~\cite{Fu2004}, Copyright 2004 Springer Nature. \textbf{b)} DFT (inset) can determine the effect of defects on positron lifetime characteristics. From~\cite{Tuomisto2020}, CC BY-NC-ND 4.0. \textbf{c)} DFT can be used to predict the radial probability distribution and compared to extended X-ray absorption fine structure experiments. From~\cite{Andrianov2021}, Copyright 2021 Elsevier. \textbf{d)} MD schemes can simulate the degradation in thermal diffusivity with dose and can be compared to TGS experiments. From~\cite{Mason2021}, CC BY 4.0. \textbf{e)} MD can also determine the system stored energy following PKA cascades and subsequent isothermal annealing. From~\cite{Hirst2021}, CC BY 4.0. \label{fig:Sim}}
\end{figure*}

The ability of PALS to detect individual vacancies allows direct comparisons to be made to primary radiation damage simulations. Tuomisto et al.~\cite{Tuomisto2020} studied Ni-ion irradiation damage in NiCoFeCr and its derivative alloys through PALS, molecular dynamics (MD), and density functional theory (DFT) simulations, shown in \cref{fig:Sim}(b). They compare the modeled fraction of vacancies as a a function of dose to their experimental data to rule out potential mechanisms and in doing so reveal the segregation of Ni and Co to vacancies. Soneda et al.~\cite{Soneda2003} use kMC simulations to model defect accumulation in Fe and demonstrate strong agreement between their predicted number densities of vacancy clusters and earlier PAS work. Further information on the use of PAS for irradiation materials can be found in a comprehensive review by Selim~\cite{Selim2021}.


\section{X-ray based methods} 

\subsection{Scattering}

X-ray scattering has been extensively used to probe radiation damage in metals due to its sensitivity to defects below the resolution limit of TEM and the ability to directly compare measurements to scattering theory. Many studies have used X-ray diffuse scattering (XDS) to characterize the size distribution of dislocation loops in irradiated Cu~\cite{Larson1975,Larson1987,Ehrhart1989}, Ni~\cite{Narayan1977,Larson1987,Ehrhart1989,Yuya1999,Olsen2016}, and W~\cite{Sun2018}. Most of these studies compare their XDS results to the size distribution of loops determined from TEM measurements. These comparisons explicitly demonstrate a) the presence of a significant fraction of loops with radii below 1~nm, and b) the inability of TEM to characterize this population of defects, an example from Olsen et al.'s work~\cite{Olsen2016} is shown in  \cref{fig:Exp}(c).

More recently, the convolutional multiple whole profile (CMWP) X-ray diffraction Line Profile Analysis (XLPA) method~\cite{Ribarik2008,Ribarik2020} has been developed to characterize dislocation loops in irradiated materials~\cite{Seymour2017,Ungar2021a}. Through calculation of the broadening of X-ray diffraction (XRD) peaks, the type, density, and size distribution of dislocation loops can be determined. Ungar et al.~\cite{Ungar2021} demonstrated this in proton-irradiated Zircaloy-2 where they combined TEM and synchrotron XRD to calculate the power law size density function of $\langle$a$\rangle$-loops, showing a significant density of loops below TEM resolution. Additionally, they extend their analysis to determine the average loop diameter as a function of loop density. They report, counter intuitively, that initially a low density of large $\langle$a$\rangle$-loops forms due to growth without impingement on other loops. Further work at low doses is needed to validate this claim.

\subsection{Absorption}

In addition to X-ray scattering, the local environment around atoms can be investigated using extended X-ray absorption fine structure (EXAFS) spectroscopy. EXAFS allows calculation of the radial distribution function which reveals the presence of defects through changes in the local coordination. This enables experiments to be directly compared to simulations, as demonstrated by Andrianov et al. in \cref{fig:Sim}(c). EXAFS has been used to study the impact of self-irradiation in Pu and its alloys as a function of dose~\cite{Booth2007,Booth2013,Olive2016,Booth2017}. Booth et al. use EXAFS to determine the fraction of a sample that is damaged and compare these measurements to estimates based on point defect production and annealing~\cite{Booth2013}. From this comparison, they report that classical defect production models (NRT-dpa and Kinchin-Pease) are insufficient to capture the effect of different bonding on defect evolution in intermetallics. Further validation of this finding with MD simulations (as demonstrated by Okamoto et al.~\cite{Okamoto1991} for N-ion irradiated amorphous Pd$_{80}$Si$_{20}$, and Booth et al.~\cite{Booth2017} for pure Pu) would be insightful, but subject to the availability of accurate interatomic potentials. 

The ability of EXAFS to probe the local structure also extends to chemical environments. Kuri et al. investigated neutron-irradiated, annealed and re-irradiated, RPV steels to analyze the coordination of solutes Cu, Ni, and Mn~\cite{Kuri2009}. They report similar atomic environments for each of the solute atoms, with attenuation in the radial distribution function peaks attributed to local disorder. The nearest neighbor coordination of Fe atoms does not decrease, which suggests the presence of vacancy-solute complexes, as has been demonstrated in PAS~\cite{Nagai2003}.

\section{Rutherford backscattering} 

Rutherford backscattering in the channeling condition (RBS/C) is part of a family of ion beam analysis techniques that have been used to characterize disorder in irradiated materials~\cite{Chu1973,Swanson1982,Matzke1985}. Placing a reference material in a channeling condition under light ion bombardment dramatically reduces the backscattered ion yield. However, by introducing defects or disorder into the lattice via irradiation, that backscattered yield increases. By varying the energy of the incident ions, the depth dependence of radiation damage can be probed. Comparing these data to the predicted damage profile allows for interpretation of the mobility of irradiation-induced defects, as shown in ref.~\cite{Lu2016}.

The radiation resistance of Ni-based complex solid solution alloys (CSSAs) has been extensively explored using RBS/C~\cite{Lu2016,Zhang2015,Jin2016,Velisa2017,Fan2019}. Through comparison of the backscattering yield of different alloys irradiated to the same dose, the effect of chemical complexity on recombination can be deduced. The morphology of larger defect clusters is found to evolve as a function of dose, with dislocation loops forming at higher dose levels, which reduce the irradiation-induced strain in the material~\cite{Jin2016,Velisa2017}.

Most studies use measures of the relative disorder between pristine and as-irradiated specimens to investigate the accumulated defects. However, further insight into nanoscale defect populations can be gained from RBS/C spectra simulated from atomistic configurations~\cite{Zhang2016,Zhang2017,Levo2021}. Zhang et al.'s simulation of ion-irradiated Ni indicate that the contribution to the RBS/C signal from extended defects is stronger than the contribution from Frenkel pairs. While this appears to conflict with previous studies which report a decrease in irradiation-induced strain and thus decrease in backscattering signal with defect agglomeration, this work demonstrates the insight that can be gained from a combination of RBS/C experiments and simulations.

\vspace{20pt}

\section{Transient grating spectroscopy} 

Recently, local thermophysical properties as measured using transient grating spectroscopy (TGS) have been applied to the measurement of radiation effects on metals and alloys~\cite{Short2015,Dennett2016,Hofmann2019}. In these experiments, a micron-scale 1D periodic laser excitation causes local heating and thermoelastic generation of surface-confined acoustic waves~\cite{Hofmann2019,Kading1995,Johnson2012}. By optically measuring the dynamics of these excitations, elastic and thermal properties can be extracted simultaneously from a single non-destructive measurement~\cite{Dennett2018a}. The measurement sampling depth is on the order of single microns, a length scale particularly relevant for the study of metals with defects induced through ion beam irradiation~\cite{Dennett2018,AlMousa2021,Hofmann2015,Hofmann2015a}. Capabilities for \emph{in~situ} TGS measurements during high temperature ion beam exposure have also been developed~\cite{Dennett2019}.

TGS has been used for nanoscale defect inference in several studies. Ferry and et al. used changes of thermal diffusivity in Si-ion irradiated Nb to infer the agglomeration of point defects into clusters, reducing total thermal carrier scattering~\cite{Ferry2019}. The most direct demonstration of nanoscale defect inference through TGS has come from Reza and coworkers~\cite{Reza2020,Reza2021}. Initially, measurements of thermal diffusivity in self-ion irradiated W specimens were used in concert with a point defect-electron scattering model to infer the population of Frenkel defects retained after room temperature irradiation to dose levels spanning five orders of magnitude~\cite{Reza2020}. Comparing with TEM-measured defect populations, a total defect density approximately an order of magnitude higher across all conditions was inferred. By using MD defect generation simulations to predict the nanoscale defect populations missed by TEM, the TGS-inferred and TEM/MD combination defect densities become comparable. In follow-on work, Reza and et al. use \emph{in~situ} TGS during thermal annealing of previously-irradiated samples to infer point defect densities and explore their recombination kinetics~\cite{Reza2021}, shown in \cref{fig:Exp}(d). They then compared their work to MD estimations of the thermal diffusivity degradation with increasing dose, shown in \cref{fig:Sim}(d).

Using local elastic properties measured with \emph{in~situ} TGS during high temperature ion irradiation, Dennett and et al. studied the accumulation of point defects, vacancy clusters, and eventually void swelling in a series of Ni-based CSSAs up to and including 5-component, single phase Cantor alloy~\cite{Dennett2021}. Alloy chemistries known to retain a higher density of nanoscale vacancy clusters, such as ternary NiCoCr, showed a much stronger elastic signature of vacancy-type defect density. Although TGS returns both elastic and thermal properties simultaneously, no work has yet demonstrated an inference method based on the combination of these properties together.

\section{Calorimetry} 
The characteristic energies of formation and migration can be used to infer the density of irradiation-induced defects through calorimetric experiments. Annealing irradiated materials leads to multiple stages of recovery~\cite{Schilling1978} where the recovery onset temperature can be used to determine the migration energy, therefore defect type, and the enthalpy change divided by the formation energy, therefore defect density. Due to their high stored energy density ($\sim$10$^3$~J/g)~\cite{Snead2019} most prior literature focuses on the recovery of irradiated ceramics, including studies on graphite~\cite{Iwata1985}, UO$_2$~\cite{Staicu2010}, SiC~\cite{Snead2019}, and NaCl~\cite{Vainshtein2000}. However, this strategy is also applicable to metals.

\begin{table*}[!ht]
\begin{tabular}{p{0.2\textwidth}>{\centering}p{0.1\textwidth}>{\centering}p{0.1\textwidth}>{\centering}p{0.1\textwidth}>{\centering}p{0.1\textwidth}>{\centering}p{0.1\textwidth}p{0.1\textwidth}<{\centering}}\hline
 & Resistivity & PAS & X-ray & RBS/C & TGS & Calorimetry \\ \hline
Max. probing range & cm & mm & cm & 10~$\mu$m & 10~$\mu$m & cm \\
Polycrystalline samples? & \ding{51} & \ding{51} & \ding{51} (XLPA) & \ding{55} & \ding{51} & \ding{51} \\ 
Point/Extended Defects & Point & Point & Both & Both & Both & Both \\
Vacancy/Interstitial & Both & Vac. & Both & Both & Both & Both\\
Chemical sensitivity? & \ding{51} & \ding{51} (CDB) & \ding{51} (EXAFS) & \ding{55}& \ding{51}& \ding{51}\\ 
Destructive? & \ding{55} & \ding{55} & \ding{55} & \ding{51} & \ding{55} & \ding{51} \\  \hline
\end{tabular}
\caption{Comparison of some salient features of the reviewed defect inference techniques. All methods listed can be applied to both pure metals as well as alloys and for irradiation with both neutrons and ions. Note: classifications should be understood as generalizations and abilities thus far demonstrated, not as absolute limitations.\label{tab:compare}}
\end{table*}

After Eugene Wigner first postulated that energy could be stored in metals as irradiation-induced defects~\cite{Wigner1942}, many works investigated defect recovery at cryogenic temperatures. These include studies on Cu~\cite{Blewitt1959,Losehand1969,Richard1990}, Al~\cite{Isebeck1966}, Mg~\cite{Delaplace1968}, and Be~\cite{Nicoud1968}. Often the focus of these works was the fundamental science behind primary radiation damage production, rather than investigating defect evolution at engineering-relevant temperatures. Research has been conducted following ambient temperature irradiation, including studies on Mo~\cite{Kinchin1958,Pedchenko1971,Lambri2009}, Cu~\cite{Blewitt1961,Pedchenko1971}, Ni and Fe~\cite{Toktogulova2010}, and Pu~\cite{Ennaceur2018}. But little work has been conducted following irradiation at elevated temperatures, with Lee et al.'s analysis of Zr-U~\cite{Lee2007} and Hirst and colleagues' recent study of Ti~\cite{Hirst2021}, shown in \cref{fig:Exp}(e), the only research to date. In addition, very few prior studies have leveraged the information contained within the fine structure of annealing spectra, with the exception of Richard et al.~\cite{Richard1990}. The authors of this study derived a kinetic model which was fit to the experimental recovery of neutron-irradiated Cu. This model yielded information on the activation energies of multiple substages and was used to (in)validate the proposed recovery mechanism for reactor-irradiated metals.

Calorimetric measurements have the key advantage of being directly comparable to simulations of radiation damage. However, until recently, this comparison had only been conducted for ceramics~\cite{Beland2013}. Hirst et al. used MD simulations to investigate the evolution of defects below the resolution limit of TEM in neutron-irradiated Ti~\cite{Hirst2021}. Their isothermal annealing simulations, shown in \cref{fig:Sim}(e), reveal a new mechanism for radiation damage recovery, with dislocation loops sweeping up point defects as they glide. This process was correlated to differential scanning calorimetry (DSC) measurements of the stored energy release between 300-480$\degree$C, where TEM micrographs show little change in the microstructure. This work reports that TEM is unable to account for two thirds of the stored energy release and demonstrates the use of DSC as a tool to characterize defects that may be hidden to microscopy techniques.

\section{Future Directions} 

\cref{tab:compare} compares some characteristic features of each of the techniques highlighted above. While great insight into the populations of nanoscale defects can be gained from these methods, their accuracy depends on the quality of data interpretation. Without the ability to directly image defects, inference-based techniques rely on theoretical models to deduce the type, size distribution, and number density of defects in materials.

One strategy that can be used to strengthen the inference model is to combine data streams from multiple methodologies. Disparate techniques often have unique sensitivities to different defect features, offering some degree of measurement orthogonality. Correlative experiments thus allow for a more comprehensive characterization of defect populations. This can be done \textit{ex~situ}, as in Meslin et al.'s study of neutron-irradiated ferritic alloys~\cite{Meslin2010}, or \textit{in~situ}, demonstrated by Jackson's simultaneous measurement of resistivity recovery and stored energy release in deuteron-irradiated metals~\cite{Jackson1980}.

Many inference-based techniques are non-destructive, which allows for both concurrent multi-modal characterization and also the ability to observe nanoscale defect generation and evolution in real-time. Combining inference-based techniques and ion-accelerators allows for significant insight, but requires detailed consideration of the experimental geometries, as highlighted by Dennett et al.~\cite{Dennett2019}. The ability to investigate the dynamics of defect accumulation and recovery \textit{in~situ} can uncover fundamental mechanisms and also accelerate nuclear materials development. It is under this promise that recent efforts have been directed in this area~\cite{Selim2021} and continued expansion of such capabilities is a necessity.

Combining experimental methods with simulated data, as shown in \cref{fig:Sim}, can also greatly benefit the interpretation of inference-based methods. This can be done directly, as in the comparison of simulated and measured atomic radial distribution functions~\cite{Andrianov2021}, or indirectly, through the measurement and estimation of thermal diffusivity degradation using TGS and MD~\cite{Reza2020,Mason2021}. The limitations of this strategy remain the considerable range of time and length scales that radiation damage encompasses. Atomistic simulations are typically restricted to timescales below microsecond and coarser methods rely on accurate parameterization. Recent advances in kMC methods~\cite{Beland2015} have demonstrated the ability to couple displacement cascades to experimental timescales and warrant further study.

In summary, the development of inference-based techniques has significantly advanced the ability to characterize the entire spectrum of defects formed in metals and alloys under irradiation. However, truly quantitative measurement of defects at the smallest scales remains a grand challenge. By designing future experimentation and simulation to function in concert, rapid progress towards this goal may be achieved. 

\begin{acknowledgments}
C.A.H acknowledges support from the National Science Foundation (NSF) Faculty Early Career Development Program (CAREER) Grant DMR-1654548. C.A.H. and C.A.D acknowledge support from the INL Laboratory Directed Research \& Development Program under U.S. Department of Energy Idaho Operations Office Contract DE-AC07-05ID14517.


\end{acknowledgments}

\bibliography{ref}

\begin{thebibliography}{116}%
\makeatletter
\providecommand \@ifxundefined [1]{%
 \@ifx{#1\undefined}
}%
\providecommand \@ifnum [1]{%
 \ifnum #1\expandafter \@firstoftwo
 \else \expandafter \@secondoftwo
 \fi
}%
\providecommand \@ifx [1]{%
 \ifx #1\expandafter \@firstoftwo
 \else \expandafter \@secondoftwo
 \fi
}%
\providecommand \natexlab [1]{#1}%
\providecommand \enquote  [1]{``#1''}%
\providecommand \bibnamefont  [1]{#1}%
\providecommand \bibfnamefont [1]{#1}%
\providecommand \citenamefont [1]{#1}%
\providecommand \href@noop [0]{\@secondoftwo}%
\providecommand \href [0]{\begingroup \@sanitize@url \@href}%
\providecommand \@href[1]{\@@startlink{#1}\@@href}%
\providecommand \@@href[1]{\endgroup#1\@@endlink}%
\providecommand \@sanitize@url [0]{\catcode `\\12\catcode `\$12\catcode
  `\&12\catcode `\#12\catcode `\^12\catcode `\_12\catcode `\%12\relax}%
\providecommand \@@startlink[1]{}%
\providecommand \@@endlink[0]{}%
\providecommand \url  [0]{\begingroup\@sanitize@url \@url }%
\providecommand \@url [1]{\endgroup\@href {#1}{\urlprefix }}%
\providecommand \urlprefix  [0]{URL }%
\providecommand \Eprint [0]{\href }%
\providecommand \doibase [0]{http://dx.doi.org/}%
\providecommand \selectlanguage [0]{\@gobble}%
\providecommand \bibinfo  [0]{\@secondoftwo}%
\providecommand \bibfield  [0]{\@secondoftwo}%
\providecommand \translation [1]{[#1]}%
\providecommand \BibitemOpen [0]{}%
\providecommand \bibitemStop [0]{}%
\providecommand \bibitemNoStop [0]{.\EOS\space}%
\providecommand \EOS [0]{\spacefactor3000\relax}%
\providecommand \BibitemShut  [1]{\csname bibitem#1\endcsname}%
\let\auto@bib@innerbib\@empty
\bibitem [{\citenamefont {Allen}\ \emph {et~al.}(2010)\citenamefont {Allen},
  \citenamefont {Busby}, \citenamefont {Meyer},\ and\ \citenamefont
  {Petti}}]{Allen2010}%
  \BibitemOpen
  \bibfield  {author} {\bibinfo {author} {\bibfnamefont {T.}~\bibnamefont
  {Allen}}, \bibinfo {author} {\bibfnamefont {J.}~\bibnamefont {Busby}},
  \bibinfo {author} {\bibfnamefont {M.}~\bibnamefont {Meyer}}, \ and\ \bibinfo
  {author} {\bibfnamefont {D.}~\bibnamefont {Petti}},\ }\bibfield  {title}
  {\enquote {\bibinfo {title} {Materials challenges for nuclear systems},}\
  }\href@noop {} {\bibfield  {journal} {\bibinfo  {journal} {Mater. Today}\
  }\textbf {\bibinfo {volume} {13}},\ \bibinfo {pages} {14--23} (\bibinfo
  {year} {2010})}\BibitemShut {NoStop}%
\bibitem [{\citenamefont {Gilbert}\ \emph {et~al.}(2021)\citenamefont
  {Gilbert}, \citenamefont {Arakawa}, \citenamefont {Bergstrom}, \citenamefont
  {Caturla}, \citenamefont {Dudarev}, \citenamefont {Gao}, \citenamefont
  {Goryaeva}, \citenamefont {Hu}, \citenamefont {Hu}, \citenamefont {Kurtz},
  \citenamefont {Litnovsky}, \citenamefont {Marian}, \citenamefont {Marinica},
  \citenamefont {Martinez}, \citenamefont {Marquis}, \citenamefont {Mason},
  \citenamefont {Nguyen}, \citenamefont {Olsson}, \citenamefont {Osetskiy},
  \citenamefont {Senor}, \citenamefont {Setyawan}, \citenamefont {Short},
  \citenamefont {Suzudo}, \citenamefont {Trelewicz}, \citenamefont {Tsuru},
  \citenamefont {Was}, \citenamefont {Wirth}, \citenamefont {Yang},
  \citenamefont {Zhang},\ and\ \citenamefont {Zinkle}}]{Gilbert2021}%
  \BibitemOpen
  \bibfield  {author} {\bibinfo {author} {\bibfnamefont {M.~R.}\ \bibnamefont
  {Gilbert}}, \bibinfo {author} {\bibfnamefont {K.}~\bibnamefont {Arakawa}},
  \bibinfo {author} {\bibfnamefont {Z.}~\bibnamefont {Bergstrom}}, \bibinfo
  {author} {\bibfnamefont {M.~J.}\ \bibnamefont {Caturla}}, \bibinfo {author}
  {\bibfnamefont {S.~L.}\ \bibnamefont {Dudarev}}, \bibinfo {author}
  {\bibfnamefont {F.}~\bibnamefont {Gao}}, \bibinfo {author} {\bibfnamefont
  {A.~M.}\ \bibnamefont {Goryaeva}}, \bibinfo {author} {\bibfnamefont {S.~Y.}\
  \bibnamefont {Hu}}, \bibinfo {author} {\bibfnamefont {X.}~\bibnamefont {Hu}},
  \bibinfo {author} {\bibfnamefont {R.~J.}\ \bibnamefont {Kurtz}}, \bibinfo
  {author} {\bibfnamefont {A.}~\bibnamefont {Litnovsky}}, \bibinfo {author}
  {\bibfnamefont {J.}~\bibnamefont {Marian}}, \bibinfo {author} {\bibfnamefont
  {M.-C.}\ \bibnamefont {Marinica}}, \bibinfo {author} {\bibfnamefont
  {E.}~\bibnamefont {Martinez}}, \bibinfo {author} {\bibfnamefont {E.~A.}\
  \bibnamefont {Marquis}}, \bibinfo {author} {\bibfnamefont {D.~R.}\
  \bibnamefont {Mason}}, \bibinfo {author} {\bibfnamefont {B.~N.}\ \bibnamefont
  {Nguyen}}, \bibinfo {author} {\bibfnamefont {P.}~\bibnamefont {Olsson}},
  \bibinfo {author} {\bibfnamefont {Y.}~\bibnamefont {Osetskiy}}, \bibinfo
  {author} {\bibfnamefont {D.}~\bibnamefont {Senor}}, \bibinfo {author}
  {\bibfnamefont {W.}~\bibnamefont {Setyawan}}, \bibinfo {author}
  {\bibfnamefont {M.~P.}\ \bibnamefont {Short}}, \bibinfo {author}
  {\bibfnamefont {T.}~\bibnamefont {Suzudo}}, \bibinfo {author} {\bibfnamefont
  {J.~R.}\ \bibnamefont {Trelewicz}}, \bibinfo {author} {\bibfnamefont
  {T.}~\bibnamefont {Tsuru}}, \bibinfo {author} {\bibfnamefont {G.~S.}\
  \bibnamefont {Was}}, \bibinfo {author} {\bibfnamefont {B.~D.}\ \bibnamefont
  {Wirth}}, \bibinfo {author} {\bibfnamefont {L.}~\bibnamefont {Yang}},
  \bibinfo {author} {\bibfnamefont {Y.}~\bibnamefont {Zhang}}, \ and\ \bibinfo
  {author} {\bibfnamefont {S.~J.}\ \bibnamefont {Zinkle}},\ }\bibfield  {title}
  {\enquote {\bibinfo {title} {Perspectives on multiscale modelling and
  experiments to accelerate materials development for fusion},}\ }\href
  {\doibase https://doi.org/10.1016/j.jnucmat.2021.153113} {\bibfield
  {journal} {\bibinfo  {journal} {J. Nucl. Mater.}\ }\textbf {\bibinfo {volume}
  {554}},\ \bibinfo {pages} {153113} (\bibinfo {year} {2021})}\BibitemShut
  {NoStop}%
\bibitem [{\citenamefont {Zinkle}\ and\ \citenamefont
  {Singh}(1993)}]{Zinkle1993}%
  \BibitemOpen
  \bibfield  {author} {\bibinfo {author} {\bibfnamefont {S.~J.}\ \bibnamefont
  {Zinkle}}\ and\ \bibinfo {author} {\bibfnamefont {B.~N.}\ \bibnamefont
  {Singh}},\ }\bibfield  {title} {\enquote {\bibinfo {title} {Analysis of
  displacement damage and defect production under cascade damage conditions},}\
  }\href {\doibase 10.1016/0022-3115(93)90140-t} {\bibfield  {journal}
  {\bibinfo  {journal} {J. Nucl. Mater.}\ }\textbf {\bibinfo {volume} {199}},\
  \bibinfo {pages} {173--191} (\bibinfo {year} {1993})}\BibitemShut {NoStop}%
\bibitem [{\citenamefont {Jenkins}\ and\ \citenamefont
  {Kirk}(2001)}]{Jenkins2001}%
  \BibitemOpen
  \bibfield  {author} {\bibinfo {author} {\bibfnamefont {M.~L.}\ \bibnamefont
  {Jenkins}}\ and\ \bibinfo {author} {\bibfnamefont {M.~A.}\ \bibnamefont
  {Kirk}},\ }\href@noop {} {\emph {\bibinfo {title} {Characterisation of
  Radiation Damage by Transmission Electron Microscopy}}}\ (\bibinfo
  {publisher} {IoP Publishing},\ \bibinfo {year} {2001})\BibitemShut {NoStop}%
\bibitem [{\citenamefont {Rickert}\ \emph {et~al.}(2022)\citenamefont
  {Rickert}, \citenamefont {Prusnick}, \citenamefont {Hunt}, \citenamefont
  {French}, \citenamefont {Turner}, \citenamefont {Dennett}, \citenamefont
  {Shao},\ and\ \citenamefont {Mann}}]{Rickert2022}%
  \BibitemOpen
  \bibfield  {author} {\bibinfo {author} {\bibfnamefont {K.}~\bibnamefont
  {Rickert}}, \bibinfo {author} {\bibfnamefont {T.~A.}\ \bibnamefont
  {Prusnick}}, \bibinfo {author} {\bibfnamefont {E.}~\bibnamefont {Hunt}},
  \bibinfo {author} {\bibfnamefont {A.}~\bibnamefont {French}}, \bibinfo
  {author} {\bibfnamefont {D.~B.}\ \bibnamefont {Turner}}, \bibinfo {author}
  {\bibfnamefont {C.~A.}\ \bibnamefont {Dennett}}, \bibinfo {author}
  {\bibfnamefont {L.}~\bibnamefont {Shao}}, \ and\ \bibinfo {author}
  {\bibfnamefont {J.~M.}\ \bibnamefont {Mann}},\ }\bibfield  {title} {\enquote
  {\bibinfo {title} {Raman and photoluminescence evaluation of ion-induced
  damage uniformity in {ThO$_2$}},}\ }\href {\doibase
  https://doi.org/10.1016/j.nimb.2022.01.011} {\bibfield  {journal} {\bibinfo
  {journal} {Nucl. Instrum. Meth. Phys. Res. B}\ }\textbf {\bibinfo {volume}
  {515}},\ \bibinfo {pages} {69--79} (\bibinfo {year} {2022})}\BibitemShut
  {NoStop}%
\bibitem [{\citenamefont {Ehrhart}(1994)}]{Ehrhart1994}%
  \BibitemOpen
  \bibfield  {author} {\bibinfo {author} {\bibfnamefont {P.}~\bibnamefont
  {Ehrhart}},\ }\bibfield  {title} {\enquote {\bibinfo {title} {Investigation
  of radiation damage by {X}-ray diffraction},}\ }\href {\doibase
  10.1016/0022-3115(94)90012-4} {\bibfield  {journal} {\bibinfo  {journal} {J.
  Nucl. Mater.}\ }\textbf {\bibinfo {volume} {216}},\ \bibinfo {pages}
  {170--198} (\bibinfo {year} {1994})}\BibitemShut {NoStop}%
\bibitem [{\citenamefont {Albertini}\ and\ \citenamefont
  {Coppola}(1992)}]{Albertini1992}%
  \BibitemOpen
  \bibfield  {author} {\bibinfo {author} {\bibfnamefont {G.}~\bibnamefont
  {Albertini}}\ and\ \bibinfo {author} {\bibfnamefont {R.}~\bibnamefont
  {Coppola}},\ }\bibfield  {title} {\enquote {\bibinfo {title} {Small-angle
  neutron scattering studies of irradiated metallic materials},}\ }\href
  {\doibase 10.1016/0168-9002(92)90979-e} {\bibfield  {journal} {\bibinfo
  {journal} {Nucl. Instrum. Methods. Phys. Res. A}\ }\textbf {\bibinfo {volume}
  {314}},\ \bibinfo {pages} {352--365} (\bibinfo {year} {1992})}\BibitemShut
  {NoStop}%
\bibitem [{\citenamefont {Swanson}(1982)}]{Swanson1982}%
  \BibitemOpen
  \bibfield  {author} {\bibinfo {author} {\bibfnamefont {M.~L.}\ \bibnamefont
  {Swanson}},\ }\bibfield  {title} {\enquote {\bibinfo {title} {The study of
  lattice defects by channelling},}\ }\href {\doibase
  10.1088/0034-4885/45/1/002} {\bibfield  {journal} {\bibinfo  {journal} {Rep.
  Prog. Phys.}\ }\textbf {\bibinfo {volume} {45}},\ \bibinfo {pages} {47--93}
  (\bibinfo {year} {1982})}\BibitemShut {NoStop}%
\bibitem [{\citenamefont {Seidman}(1978)}]{Seidman1978}%
  \BibitemOpen
  \bibfield  {author} {\bibinfo {author} {\bibfnamefont {D.~N.}\ \bibnamefont
  {Seidman}},\ }\bibfield  {title} {\enquote {\bibinfo {title} {The study of
  radiation damage in metals with the field-ion and atom-probe microscopes},}\
  }\href {\doibase 10.1016/0039-6028(78)90430-2} {\bibfield  {journal}
  {\bibinfo  {journal} {Surf. Sci.}\ }\textbf {\bibinfo {volume} {70}},\
  \bibinfo {pages} {532--565} (\bibinfo {year} {1978})}\BibitemShut {NoStop}%
\bibitem [{\citenamefont {Selim}(2021)}]{Selim2021}%
  \BibitemOpen
  \bibfield  {author} {\bibinfo {author} {\bibfnamefont {F.~A.}\ \bibnamefont
  {Selim}},\ }\bibfield  {title} {\enquote {\bibinfo {title} {Positron
  annihilation spectroscopy of defects in nuclear and irradiated materials -
  {a} review},}\ }\href {\doibase 10.1016/j.matchar.2021.110952} {\bibfield
  {journal} {\bibinfo  {journal} {Mat. Char.}\ }\textbf {\bibinfo {volume}
  {174}},\ \bibinfo {pages} {110952} (\bibinfo {year} {2021})}\BibitemShut
  {NoStop}%
\bibitem [{\citenamefont {Jenkins}(1994)}]{Jenkins1994}%
  \BibitemOpen
  \bibfield  {author} {\bibinfo {author} {\bibfnamefont {M.~L.}\ \bibnamefont
  {Jenkins}},\ }\bibfield  {title} {\enquote {\bibinfo {title}
  {Characterisation of radiation-damage microstructures by {TEM}},}\ }\href
  {\doibase 10.1016/0022-3115(94)90010-8} {\bibfield  {journal} {\bibinfo
  {journal} {J. Nucl. Mater.}\ }\textbf {\bibinfo {volume} {216}},\ \bibinfo
  {pages} {124--156} (\bibinfo {year} {1994})}\BibitemShut {NoStop}%
\bibitem [{\citenamefont {Zhou}\ \emph {et~al.}(2007)\citenamefont {Zhou},
  \citenamefont {Dudarev}, \citenamefont {Jenkins}, \citenamefont {Sutton},\
  and\ \citenamefont {Kirk}}]{Zhou2007}%
  \BibitemOpen
  \bibfield  {author} {\bibinfo {author} {\bibfnamefont {Z.}~\bibnamefont
  {Zhou}}, \bibinfo {author} {\bibfnamefont {S.~L.}\ \bibnamefont {Dudarev}},
  \bibinfo {author} {\bibfnamefont {M.~L.}\ \bibnamefont {Jenkins}}, \bibinfo
  {author} {\bibfnamefont {A.~P.}\ \bibnamefont {Sutton}}, \ and\ \bibinfo
  {author} {\bibfnamefont {M.~A.}\ \bibnamefont {Kirk}},\ }\bibfield  {title}
  {\enquote {\bibinfo {title} {Diffraction imaging and diffuse scattering by
  small dislocation loops},}\ }\href {\doibase 10.1016/j.jnucmat.2007.03.135}
  {\bibfield  {journal} {\bibinfo  {journal} {J. Nucl. Mater.}\ }\textbf
  {\bibinfo {volume} {367-370}},\ \bibinfo {pages} {305--310} (\bibinfo {year}
  {2007})}\BibitemShut {NoStop}%
\bibitem [{\citenamefont {Yi}\ \emph {et~al.}(2015)\citenamefont {Yi},
  \citenamefont {Sand}, \citenamefont {Mason}, \citenamefont {Kirk},
  \citenamefont {Roberts}, \citenamefont {Nordlund},\ and\ \citenamefont
  {Dudarev}}]{Yi2015}%
  \BibitemOpen
  \bibfield  {author} {\bibinfo {author} {\bibfnamefont {X.}~\bibnamefont
  {Yi}}, \bibinfo {author} {\bibfnamefont {A.~E.}\ \bibnamefont {Sand}},
  \bibinfo {author} {\bibfnamefont {D.~R.}\ \bibnamefont {Mason}}, \bibinfo
  {author} {\bibfnamefont {M.~A.}\ \bibnamefont {Kirk}}, \bibinfo {author}
  {\bibfnamefont {S.~G.}\ \bibnamefont {Roberts}}, \bibinfo {author}
  {\bibfnamefont {K.}~\bibnamefont {Nordlund}}, \ and\ \bibinfo {author}
  {\bibfnamefont {S.~L.}\ \bibnamefont {Dudarev}},\ }\bibfield  {title}
  {\enquote {\bibinfo {title} {Direct observation of size scaling and elastic
  interaction between nano-scale defects in collision cascades},}\ }\href@noop
  {} {\bibfield  {journal} {\bibinfo  {journal} {Euro. Phys. Lett.}\ }\textbf
  {\bibinfo {volume} {110}},\ \bibinfo {pages} {36001} (\bibinfo {year}
  {2015})}\BibitemShut {NoStop}%
\bibitem [{\citenamefont {Caturla}\ \emph {et~al.}(2000)\citenamefont
  {Caturla}, \citenamefont {Soneda}, \citenamefont {Alonso}, \citenamefont
  {Wirth}, \citenamefont {de~la Rubia},\ and\ \citenamefont
  {Perlado}}]{Caturla2000}%
  \BibitemOpen
  \bibfield  {author} {\bibinfo {author} {\bibfnamefont {M.~J.}\ \bibnamefont
  {Caturla}}, \bibinfo {author} {\bibfnamefont {N.}~\bibnamefont {Soneda}},
  \bibinfo {author} {\bibfnamefont {E.}~\bibnamefont {Alonso}}, \bibinfo
  {author} {\bibfnamefont {B.~D.}\ \bibnamefont {Wirth}}, \bibinfo {author}
  {\bibfnamefont {T.~Diaz}\ \bibnamefont {de~la Rubia}}, \ and\ \bibinfo
  {author} {\bibfnamefont {J.~M.}\ \bibnamefont {Perlado}},\ }\bibfield
  {title} {\enquote {\bibinfo {title} {Comparative study of radiation damage
  accumulation in {Cu} and {Fe}},}\ }\href {\doibase
  10.1016/s0022-3115(99)00220-2} {\bibfield  {journal} {\bibinfo  {journal} {J.
  Nucl. Mater.}\ }\textbf {\bibinfo {volume} {276}},\ \bibinfo {pages} {13--21}
  (\bibinfo {year} {2000})}\BibitemShut {NoStop}%
\bibitem [{\citenamefont {Meslin}\ \emph {et~al.}(2010)\citenamefont {Meslin},
  \citenamefont {Lambrecht}, \citenamefont {Hernández-Mayoral}, \citenamefont
  {Bergner}, \citenamefont {Malerba}, \citenamefont {Pareige}, \citenamefont
  {Radiguet}, \citenamefont {Barbu}, \citenamefont {Gómez-Briceño},
  \citenamefont {Ulbricht},\ and\ \citenamefont {Almazouzi}}]{Meslin2010}%
  \BibitemOpen
  \bibfield  {author} {\bibinfo {author} {\bibfnamefont {E.}~\bibnamefont
  {Meslin}}, \bibinfo {author} {\bibfnamefont {M.}~\bibnamefont {Lambrecht}},
  \bibinfo {author} {\bibfnamefont {M.}~\bibnamefont {Hernández-Mayoral}},
  \bibinfo {author} {\bibfnamefont {F.}~\bibnamefont {Bergner}}, \bibinfo
  {author} {\bibfnamefont {L.}~\bibnamefont {Malerba}}, \bibinfo {author}
  {\bibfnamefont {P.}~\bibnamefont {Pareige}}, \bibinfo {author} {\bibfnamefont
  {B.}~\bibnamefont {Radiguet}}, \bibinfo {author} {\bibfnamefont
  {A.}~\bibnamefont {Barbu}}, \bibinfo {author} {\bibfnamefont
  {D.}~\bibnamefont {Gómez-Briceño}}, \bibinfo {author} {\bibfnamefont
  {A.}~\bibnamefont {Ulbricht}}, \ and\ \bibinfo {author} {\bibfnamefont
  {A.}~\bibnamefont {Almazouzi}},\ }\bibfield  {title} {\enquote {\bibinfo
  {title} {Characterization of neutron-irradiated ferritic model alloys and a
  {RPV} steel from combined {APT}, {SANS}, {TEM} and {PAS} analyses},}\ }\href
  {\doibase 10.1016/j.jnucmat.2009.12.021} {\bibfield  {journal} {\bibinfo
  {journal} {J. Nucl. Mater.}\ }\textbf {\bibinfo {volume} {406}},\ \bibinfo
  {pages} {73--83} (\bibinfo {year} {2010})}\BibitemShut {NoStop}%
\bibitem [{\citenamefont {Reza}\ \emph {et~al.}(2020)\citenamefont {Reza},
  \citenamefont {Yu}, \citenamefont {Mizohata},\ and\ \citenamefont
  {Hofmann}}]{Reza2020}%
  \BibitemOpen
  \bibfield  {author} {\bibinfo {author} {\bibfnamefont {A.}~\bibnamefont
  {Reza}}, \bibinfo {author} {\bibfnamefont {H.}~\bibnamefont {Yu}}, \bibinfo
  {author} {\bibfnamefont {K.}~\bibnamefont {Mizohata}}, \ and\ \bibinfo
  {author} {\bibfnamefont {F.}~\bibnamefont {Hofmann}},\ }\bibfield  {title}
  {\enquote {\bibinfo {title} {Thermal diffusivity degradation and point defect
  density in self-ion implanted tungsten},}\ }\href {\doibase
  https://doi.org/10.1016/j.actamat.2020.03.034} {\bibfield  {journal}
  {\bibinfo  {journal} {Acta Mater.}\ }\textbf {\bibinfo {volume} {193}},\
  \bibinfo {pages} {270--279} (\bibinfo {year} {2020})}\BibitemShut {NoStop}%
\bibitem [{\citenamefont {Ungár}\ \emph
  {et~al.}(2021{\natexlab{a}})\citenamefont {Ungár}, \citenamefont {Frankel},
  \citenamefont {Ribárik}, \citenamefont {Race},\ and\ \citenamefont
  {Preuss}}]{Ungar2021}%
  \BibitemOpen
  \bibfield  {author} {\bibinfo {author} {\bibfnamefont {T.}~\bibnamefont
  {Ungár}}, \bibinfo {author} {\bibfnamefont {P.}~\bibnamefont {Frankel}},
  \bibinfo {author} {\bibfnamefont {G.}~\bibnamefont {Ribárik}}, \bibinfo
  {author} {\bibfnamefont {C.~P.}\ \bibnamefont {Race}}, \ and\ \bibinfo
  {author} {\bibfnamefont {M.}~\bibnamefont {Preuss}},\ }\bibfield  {title}
  {\enquote {\bibinfo {title} {Size-distribution of irradiation-induced
  dislocation-loops in materials used in the nuclear industry},}\ }\href
  {\doibase 10.1016/j.jnucmat.2021.152945} {\bibfield  {journal} {\bibinfo
  {journal} {J. Nucl. Mater.}\ }\textbf {\bibinfo {volume} {550}},\ \bibinfo
  {pages} {152945} (\bibinfo {year} {2021}{\natexlab{a}})}\BibitemShut
  {NoStop}%
\bibitem [{\citenamefont {Li}\ \emph {et~al.}(2022)\citenamefont {Li},
  \citenamefont {Beyerlein},\ and\ \citenamefont {Han}}]{Li2022}%
  \BibitemOpen
  \bibfield  {author} {\bibinfo {author} {\bibfnamefont {J-T.}\ \bibnamefont
  {Li}}, \bibinfo {author} {\bibfnamefont {I.~J.}\ \bibnamefont {Beyerlein}}, \
  and\ \bibinfo {author} {\bibfnamefont {W-Z.}\ \bibnamefont {Han}},\
  }\bibfield  {title} {\enquote {\bibinfo {title} {Helium irradiation-induced
  ultrahigh hardening in niobium},}\ }\href {\doibase
  10.1016/j.actamat.2022.117656} {\bibfield  {journal} {\bibinfo  {journal}
  {Acta Mater.}\ }\textbf {\bibinfo {volume} {226}},\ \bibinfo {pages} {117656}
  (\bibinfo {year} {2022})}\BibitemShut {NoStop}%
\bibitem [{\citenamefont {Shelyug}\ \emph {et~al.}(2018)\citenamefont
  {Shelyug}, \citenamefont {Palomares}, \citenamefont {Lang},\ and\
  \citenamefont {Navrotsky}}]{Shelyug2018}%
  \BibitemOpen
  \bibfield  {author} {\bibinfo {author} {\bibfnamefont {A.}~\bibnamefont
  {Shelyug}}, \bibinfo {author} {\bibfnamefont {R.~I.}\ \bibnamefont
  {Palomares}}, \bibinfo {author} {\bibfnamefont {M.}~\bibnamefont {Lang}}, \
  and\ \bibinfo {author} {\bibfnamefont {A.}~\bibnamefont {Navrotsky}},\
  }\bibfield  {title} {\enquote {\bibinfo {title} {Energetics of defect
  production in fluorite-structured {C}e{O}$_2$ induced by highly ionizing
  radiation},}\ }\href {\doibase 10.1103/PhysRevMaterials.2.093607} {\bibfield
  {journal} {\bibinfo  {journal} {Phys. Rev. Mat.}\ }\textbf {\bibinfo {volume}
  {2}},\ \bibinfo {pages} {093607} (\bibinfo {year} {2018})}\BibitemShut
  {NoStop}%
\bibitem [{\citenamefont {Khanolkar}\ \emph {et~al.}(2022)\citenamefont
  {Khanolkar}, \citenamefont {Dennett}, \citenamefont {Hua}, \citenamefont
  {Mann}, \citenamefont {Hurley},\ and\ \citenamefont
  {Khafizov}}]{Khanolkar2022}%
  \BibitemOpen
  \bibfield  {author} {\bibinfo {author} {\bibfnamefont {A.}~\bibnamefont
  {Khanolkar}}, \bibinfo {author} {\bibfnamefont {C.~A.}\ \bibnamefont
  {Dennett}}, \bibinfo {author} {\bibfnamefont {Z.}~\bibnamefont {Hua}},
  \bibinfo {author} {\bibfnamefont {J.~M.}\ \bibnamefont {Mann}}, \bibinfo
  {author} {\bibfnamefont {D.~H.}\ \bibnamefont {Hurley}}, \ and\ \bibinfo
  {author} {\bibfnamefont {M.}~\bibnamefont {Khafizov}},\ }\bibfield  {title}
  {\enquote {\bibinfo {title} {Inferring relative dose-dependent color center
  populations in proton irradiated thoria single crystals using optical
  spectroscopy},}\ }\href@noop {} {\bibfield  {journal} {\bibinfo  {journal}
  {Phys. Chem. Chem. Phys.}\ }\textbf {\bibinfo {volume} {24}},\ \bibinfo
  {pages} {6133--6145} (\bibinfo {year} {2022})}\BibitemShut {NoStop}%
\bibitem [{\citenamefont {Briggmann}\ \emph {et~al.}(1994)\citenamefont
  {Briggmann}, \citenamefont {Dagge}, \citenamefont {Frank}, \citenamefont
  {Seeger}, \citenamefont {Stoll},\ and\ \citenamefont
  {Verbruggen}}]{Briggmann1994}%
  \BibitemOpen
  \bibfield  {author} {\bibinfo {author} {\bibfnamefont {J.}~\bibnamefont
  {Briggmann}}, \bibinfo {author} {\bibfnamefont {K.}~\bibnamefont {Dagge}},
  \bibinfo {author} {\bibfnamefont {W.}~\bibnamefont {Frank}}, \bibinfo
  {author} {\bibfnamefont {A.}~\bibnamefont {Seeger}}, \bibinfo {author}
  {\bibfnamefont {H.}~\bibnamefont {Stoll}}, \ and\ \bibinfo {author}
  {\bibfnamefont {A.~H.}\ \bibnamefont {Verbruggen}},\ }\bibfield  {title}
  {\enquote {\bibinfo {title} {Irradiation-induced defects in thin aluminium
  films studied by $1/f$ noise},}\ }\href {\doibase 10.1002/pssa.2211460128}
  {\bibfield  {journal} {\bibinfo  {journal} {Phys. Status Solidi A}\ }\textbf
  {\bibinfo {volume} {146}},\ \bibinfo {pages} {325--335} (\bibinfo {year}
  {1994})}\BibitemShut {NoStop}%
\bibitem [{\citenamefont {Agarwal}\ \emph {et~al.}(2020)\citenamefont
  {Agarwal}, \citenamefont {Liedke}, \citenamefont {Jones}, \citenamefont
  {Reed}, \citenamefont {Kohnert}, \citenamefont {Uberuaga}, \citenamefont
  {Wang}, \citenamefont {Cooper}, \citenamefont {Kaoumi}, \citenamefont {Li},
  \citenamefont {Auguste}, \citenamefont {Hosemann}, \citenamefont {Capolungo},
  \citenamefont {Edwards}, \citenamefont {Butterling}, \citenamefont
  {Hirschmann}, \citenamefont {Wagner},\ and\ \citenamefont
  {Selim}}]{Agarwal2020}%
  \BibitemOpen
  \bibfield  {author} {\bibinfo {author} {\bibfnamefont {S.}~\bibnamefont
  {Agarwal}}, \bibinfo {author} {\bibfnamefont {M.~O.}\ \bibnamefont {Liedke}},
  \bibinfo {author} {\bibfnamefont {A.~C.~L.}\ \bibnamefont {Jones}}, \bibinfo
  {author} {\bibfnamefont {E.}~\bibnamefont {Reed}}, \bibinfo {author}
  {\bibfnamefont {A.~A.}\ \bibnamefont {Kohnert}}, \bibinfo {author}
  {\bibfnamefont {B.~P.}\ \bibnamefont {Uberuaga}}, \bibinfo {author}
  {\bibfnamefont {Y.~Q.}\ \bibnamefont {Wang}}, \bibinfo {author}
  {\bibfnamefont {J.}~\bibnamefont {Cooper}}, \bibinfo {author} {\bibfnamefont
  {D.}~\bibnamefont {Kaoumi}}, \bibinfo {author} {\bibfnamefont
  {N.}~\bibnamefont {Li}}, \bibinfo {author} {\bibfnamefont {R.}~\bibnamefont
  {Auguste}}, \bibinfo {author} {\bibfnamefont {P.}~\bibnamefont {Hosemann}},
  \bibinfo {author} {\bibfnamefont {L.}~\bibnamefont {Capolungo}}, \bibinfo
  {author} {\bibfnamefont {D.~J.}\ \bibnamefont {Edwards}}, \bibinfo {author}
  {\bibfnamefont {M.}~\bibnamefont {Butterling}}, \bibinfo {author}
  {\bibfnamefont {E.}~\bibnamefont {Hirschmann}}, \bibinfo {author}
  {\bibfnamefont {A.}~\bibnamefont {Wagner}}, \ and\ \bibinfo {author}
  {\bibfnamefont {F.~A.}\ \bibnamefont {Selim}},\ }\bibfield  {title} {\enquote
  {\bibinfo {title} {A new mechanism for void-cascade interaction from
  nondestructive depth-resolved atomic-scale measurements of ion
  irradiation–induced defects in {Fe}},}\ }\href@noop {} {\bibfield
  {journal} {\bibinfo  {journal} {Sci. Adv.}\ }\textbf {\bibinfo {volume} {6}}
  (\bibinfo {year} {2020})}\BibitemShut {NoStop}%
\bibitem [{\citenamefont {Olsen}\ \emph {et~al.}(2016)\citenamefont {Olsen},
  \citenamefont {Jin}, \citenamefont {Lu}, \citenamefont {Beland},
  \citenamefont {Wang}, \citenamefont {Bei}, \citenamefont {Specht},\ and\
  \citenamefont {Larson}}]{Olsen2016}%
  \BibitemOpen
  \bibfield  {author} {\bibinfo {author} {\bibfnamefont {R.~J.}\ \bibnamefont
  {Olsen}}, \bibinfo {author} {\bibfnamefont {K.}~\bibnamefont {Jin}}, \bibinfo
  {author} {\bibfnamefont {C.}~\bibnamefont {Lu}}, \bibinfo {author}
  {\bibfnamefont {L.~K.}\ \bibnamefont {Beland}}, \bibinfo {author}
  {\bibfnamefont {L.}~\bibnamefont {Wang}}, \bibinfo {author} {\bibfnamefont
  {H.}~\bibnamefont {Bei}}, \bibinfo {author} {\bibfnamefont {E.~D.}\
  \bibnamefont {Specht}}, \ and\ \bibinfo {author} {\bibfnamefont {B.~C.}\
  \bibnamefont {Larson}},\ }\bibfield  {title} {\enquote {\bibinfo {title}
  {Investigation of defect clusters in ion-irradiated {Ni} and {NiCo} using
  diffuse {X}-ray scattering and electron microscopy},}\ }\href {\doibase
  10.1016/j.jnucmat.2015.11.030} {\bibfield  {journal} {\bibinfo  {journal} {J.
  Nucl. Mater.}\ }\textbf {\bibinfo {volume} {469}},\ \bibinfo {pages}
  {153--161} (\bibinfo {year} {2016})}\BibitemShut {NoStop}%
\bibitem [{\citenamefont {Reza}\ \emph {et~al.}(2022)\citenamefont {Reza},
  \citenamefont {He}, \citenamefont {Dennett}, \citenamefont {Yu},
  \citenamefont {Mizohata},\ and\ \citenamefont {Hofmann}}]{Reza2021}%
  \BibitemOpen
  \bibfield  {author} {\bibinfo {author} {\bibfnamefont {A.}~\bibnamefont
  {Reza}}, \bibinfo {author} {\bibfnamefont {G.}~\bibnamefont {He}}, \bibinfo
  {author} {\bibfnamefont {C.~A.}\ \bibnamefont {Dennett}}, \bibinfo {author}
  {\bibfnamefont {H.}~\bibnamefont {Yu}}, \bibinfo {author} {\bibfnamefont
  {K.}~\bibnamefont {Mizohata}}, \ and\ \bibinfo {author} {\bibfnamefont
  {F.}~\bibnamefont {Hofmann}},\ }\bibfield  {title} {\enquote {\bibinfo
  {title} {Thermal diffusivity recovery and defect annealing kinetics of
  self-ion implanted tungsten probed by \emph{in~situ} transient grating
  spectroscopy},}\ }\href@noop {} {\bibfield  {journal} {\bibinfo  {journal}
  {Acta Mater.}\ }\textbf {\bibinfo {volume} {232}},\ \bibinfo {pages} {117926}
  (\bibinfo {year} {2022})}\BibitemShut {NoStop}%
\bibitem [{\citenamefont {Hirst}\ \emph {et~al.}(2021)\citenamefont {Hirst},
  \citenamefont {Granberg}, \citenamefont {Kombaiah}, \citenamefont {Cao},
  \citenamefont {Middlemas}, \citenamefont {Kemp}, \citenamefont {Li},
  \citenamefont {Nordlund},\ and\ \citenamefont {Short}}]{Hirst2021}%
  \BibitemOpen
  \bibfield  {author} {\bibinfo {author} {\bibfnamefont {C.~A.}\ \bibnamefont
  {Hirst}}, \bibinfo {author} {\bibfnamefont {F.}~\bibnamefont {Granberg}},
  \bibinfo {author} {\bibfnamefont {B.}~\bibnamefont {Kombaiah}}, \bibinfo
  {author} {\bibfnamefont {P.}~\bibnamefont {Cao}}, \bibinfo {author}
  {\bibfnamefont {S.}~\bibnamefont {Middlemas}}, \bibinfo {author}
  {\bibfnamefont {R.~S.}\ \bibnamefont {Kemp}}, \bibinfo {author}
  {\bibfnamefont {J.}~\bibnamefont {Li}}, \bibinfo {author} {\bibfnamefont
  {K.}~\bibnamefont {Nordlund}}, \ and\ \bibinfo {author} {\bibfnamefont
  {M.~P.}\ \bibnamefont {Short}},\ }\bibfield  {title} {\enquote {\bibinfo
  {title} {Revealing hidden defects through stored energy measurements of
  radiation damage},}\ }\href@noop {} {\  (\bibinfo {year} {2021})},\ \bibinfo
  {note} {https://arxiv.org/abs/2109.15138}\BibitemShut {NoStop}%
\bibitem [{\citenamefont {Broom}(1954)}]{Broom1954}%
  \BibitemOpen
  \bibfield  {author} {\bibinfo {author} {\bibfnamefont {T.}~\bibnamefont
  {Broom}},\ }\bibfield  {title} {\enquote {\bibinfo {title} {Lattice defects
  and the electrical resistivity of metals},}\ }\href {\doibase
  10.1080/00018735400101163} {\bibfield  {journal} {\bibinfo  {journal} {Adv.
  Phys.}\ }\textbf {\bibinfo {volume} {3}},\ \bibinfo {pages} {26--83}
  (\bibinfo {year} {1954})}\BibitemShut {NoStop}%
\bibitem [{\citenamefont {Matthiessen}\ and\ \citenamefont
  {Vogt}(1864)}]{Matthiessen1864}%
  \BibitemOpen
  \bibfield  {author} {\bibinfo {author} {\bibfnamefont {A.}~\bibnamefont
  {Matthiessen}}\ and\ \bibinfo {author} {\bibfnamefont {C.}~\bibnamefont
  {Vogt}},\ }\bibfield  {title} {\enquote {\bibinfo {title} {Ueber den einfluss
  der temperatur auf die elektrische leitungsfaehigkeit der legirungen},}\
  }\href {\doibase 10.1002/andp.18641980504} {\bibfield  {journal} {\bibinfo
  {journal} {Adv. Phys.}\ }\textbf {\bibinfo {volume} {198}},\ \bibinfo {pages}
  {19--78} (\bibinfo {year} {1864})}\BibitemShut {NoStop}%
\bibitem [{\citenamefont {Lucasson}\ and\ \citenamefont
  {Walker}(1962)}]{Lucasson1962}%
  \BibitemOpen
  \bibfield  {author} {\bibinfo {author} {\bibfnamefont {P.~G.}\ \bibnamefont
  {Lucasson}}\ and\ \bibinfo {author} {\bibfnamefont {R.~M.}\ \bibnamefont
  {Walker}},\ }\bibfield  {title} {\enquote {\bibinfo {title} {Production and
  recovery of electron-induced radiation damage in a number of metals},}\
  }\href {\doibase 10.1103/physrev.127.485} {\bibfield  {journal} {\bibinfo
  {journal} {Phys. Rev.}\ }\textbf {\bibinfo {volume} {127}},\ \bibinfo {pages}
  {485--500} (\bibinfo {year} {1962})}\BibitemShut {NoStop}%
\bibitem [{\citenamefont {Nakagawa}\ \emph {et~al.}(1977)\citenamefont
  {Nakagawa}, \citenamefont {Boening}, \citenamefont {Rosner},\ and\
  \citenamefont {Vogl}}]{Nakagawa1977}%
  \BibitemOpen
  \bibfield  {author} {\bibinfo {author} {\bibfnamefont {M.}~\bibnamefont
  {Nakagawa}}, \bibinfo {author} {\bibfnamefont {K.}~\bibnamefont {Boening}},
  \bibinfo {author} {\bibfnamefont {P.}~\bibnamefont {Rosner}}, \ and\ \bibinfo
  {author} {\bibfnamefont {G.}~\bibnamefont {Vogl}},\ }\bibfield  {title}
  {\enquote {\bibinfo {title} {High-dose neutron-irradiation effects in fcc
  metals at 4.6 {K}},}\ }\href {\doibase 10.1103/physrevb.16.5285} {\bibfield
  {journal} {\bibinfo  {journal} {Phys. Rev. B}\ }\textbf {\bibinfo {volume}
  {16}},\ \bibinfo {pages} {5285--5302} (\bibinfo {year} {1977})}\BibitemShut
  {NoStop}%
\bibitem [{\citenamefont {Nakagawa}\ \emph {et~al.}(1979)\citenamefont
  {Nakagawa}, \citenamefont {Mansel}, \citenamefont {Boening}, \citenamefont
  {Rosner},\ and\ \citenamefont {Vogl}}]{Nakagawa1979}%
  \BibitemOpen
  \bibfield  {author} {\bibinfo {author} {\bibfnamefont {M.}~\bibnamefont
  {Nakagawa}}, \bibinfo {author} {\bibfnamefont {W.}~\bibnamefont {Mansel}},
  \bibinfo {author} {\bibfnamefont {K.}~\bibnamefont {Boening}}, \bibinfo
  {author} {\bibfnamefont {P.}~\bibnamefont {Rosner}}, \ and\ \bibinfo {author}
  {\bibfnamefont {G.}~\bibnamefont {Vogl}},\ }\bibfield  {title} {\enquote
  {\bibinfo {title} {Spontaneous recombination volumes of {F}renkel defects in
  neutron-irradiated non-fcc metals},}\ }\href {\doibase
  10.1103/physrevb.19.742} {\bibfield  {journal} {\bibinfo  {journal} {Phys.
  Rev. B}\ }\textbf {\bibinfo {volume} {19}},\ \bibinfo {pages} {742--748}
  (\bibinfo {year} {1979})}\BibitemShut {NoStop}%
\bibitem [{\citenamefont {Nakagawa}(1982)}]{Nakagawa1982}%
  \BibitemOpen
  \bibfield  {author} {\bibinfo {author} {\bibfnamefont {M.}~\bibnamefont
  {Nakagawa}},\ }\bibfield  {title} {\enquote {\bibinfo {title} {Saturation
  phenomena in irradiated metals at low temperature},}\ }\href {\doibase
  10.1016/0022-3115(82)90487-1} {\bibfield  {journal} {\bibinfo  {journal} {J.
  Nucl. Mater.}\ }\textbf {\bibinfo {volume} {108-109}},\ \bibinfo {pages}
  {194--200} (\bibinfo {year} {1982})}\BibitemShut {NoStop}%
\bibitem [{\citenamefont {Iwase}\ \emph {et~al.}(1992)\citenamefont {Iwase},
  \citenamefont {Iwata}, \citenamefont {Nihira},\ and\ \citenamefont
  {Sasaki}}]{Iwase1992}%
  \BibitemOpen
  \bibfield  {author} {\bibinfo {author} {\bibfnamefont {A.}~\bibnamefont
  {Iwase}}, \bibinfo {author} {\bibfnamefont {T.}~\bibnamefont {Iwata}},
  \bibinfo {author} {\bibfnamefont {T.}~\bibnamefont {Nihira}}, \ and\ \bibinfo
  {author} {\bibfnamefont {S.}~\bibnamefont {Sasaki}},\ }\bibfield  {title}
  {\enquote {\bibinfo {title} {Defect recovery and radiation annealing in fcc
  metals irradiated with high energy ions},}\ }\href {\doibase
  10.1080/10420159208219833} {\bibfield  {journal} {\bibinfo  {journal}
  {Radiat. Eff. Defects Solids}\ }\textbf {\bibinfo {volume} {124}},\ \bibinfo
  {pages} {117--126} (\bibinfo {year} {1992})}\BibitemShut {NoStop}%
\bibitem [{\citenamefont {Chimi}\ \emph {et~al.}(2000)\citenamefont {Chimi},
  \citenamefont {Iwase}, \citenamefont {Ishikawa}, \citenamefont {Kuroda},\
  and\ \citenamefont {Kambara}}]{Chimi2000}%
  \BibitemOpen
  \bibfield  {author} {\bibinfo {author} {\bibfnamefont {Y.}~\bibnamefont
  {Chimi}}, \bibinfo {author} {\bibfnamefont {A.}~\bibnamefont {Iwase}},
  \bibinfo {author} {\bibfnamefont {N.}~\bibnamefont {Ishikawa}}, \bibinfo
  {author} {\bibfnamefont {N.}~\bibnamefont {Kuroda}}, \ and\ \bibinfo {author}
  {\bibfnamefont {T.}~\bibnamefont {Kambara}},\ }\bibfield  {title} {\enquote
  {\bibinfo {title} {Radiation annealing induced by electronic excitation in
  iron},}\ }\href {\doibase 10.1016/s0168-583x(99)01139-8} {\bibfield
  {journal} {\bibinfo  {journal} {Nucl. Instrum. Methods. Phys. Res. B}\
  }\textbf {\bibinfo {volume} {164--165}},\ \bibinfo {pages} {408--414}
  (\bibinfo {year} {2000})}\BibitemShut {NoStop}%
\bibitem [{\citenamefont {Iwamoto}\ \emph {et~al.}(2015)\citenamefont
  {Iwamoto}, \citenamefont {Yoshiie}, \citenamefont {Yoshida}, \citenamefont
  {Nakamoto}, \citenamefont {Sakamoto}, \citenamefont {Kuriyama}, \citenamefont
  {Uesugi}, \citenamefont {Ishi}, \citenamefont {Xu}, \citenamefont {Yashima},
  \citenamefont {F.Takahashi}, \citenamefont {Mori},\ and\ \citenamefont
  {Ogitsu}}]{Iwamoto2015}%
  \BibitemOpen
  \bibfield  {author} {\bibinfo {author} {\bibfnamefont {Y.}~\bibnamefont
  {Iwamoto}}, \bibinfo {author} {\bibfnamefont {T.}~\bibnamefont {Yoshiie}},
  \bibinfo {author} {\bibfnamefont {M.}~\bibnamefont {Yoshida}}, \bibinfo
  {author} {\bibfnamefont {T.}~\bibnamefont {Nakamoto}}, \bibinfo {author}
  {\bibfnamefont {M.}~\bibnamefont {Sakamoto}}, \bibinfo {author}
  {\bibfnamefont {Y.}~\bibnamefont {Kuriyama}}, \bibinfo {author}
  {\bibfnamefont {T.}~\bibnamefont {Uesugi}}, \bibinfo {author} {\bibfnamefont
  {Y.}~\bibnamefont {Ishi}}, \bibinfo {author} {\bibfnamefont {Q.}~\bibnamefont
  {Xu}}, \bibinfo {author} {\bibfnamefont {H.}~\bibnamefont {Yashima}},
  \bibinfo {author} {\bibnamefont {F.Takahashi}}, \bibinfo {author}
  {\bibfnamefont {Y.}~\bibnamefont {Mori}}, \ and\ \bibinfo {author}
  {\bibfnamefont {T.}~\bibnamefont {Ogitsu}},\ }\bibfield  {title} {\enquote
  {\bibinfo {title} {Measurement of the displacement cross-section of copper
  irradiated with 125 {MeV} protons at 12 {K}},}\ }\href {\doibase
  10.1016/j.jnucmat.2014.12.125} {\bibfield  {journal} {\bibinfo  {journal} {J.
  Nucl. Mater.}\ }\textbf {\bibinfo {volume} {458}},\ \bibinfo {pages}
  {369--375} (\bibinfo {year} {2015})}\BibitemShut {NoStop}%
\bibitem [{\citenamefont {Horak}\ and\ \citenamefont
  {Blewitt}(1975)}]{Horak1975}%
  \BibitemOpen
  \bibfield  {author} {\bibinfo {author} {\bibfnamefont {J.~A.}\ \bibnamefont
  {Horak}}\ and\ \bibinfo {author} {\bibfnamefont {T.~H.}\ \bibnamefont
  {Blewitt}},\ }\bibfield  {title} {\enquote {\bibinfo {title} {Fast- and
  thermal-neutron irradiation and annealing of {Cu}, {Ni}, {Fe}, {Ti}, {Pd}},}\
  }\href@noop {} {\bibfield  {journal} {\bibinfo  {journal} {Nucl. Technol.}\ }
  (\bibinfo {year} {1975})}\BibitemShut {NoStop}%
\bibitem [{\citenamefont {Fu}\ \emph {et~al.}(2004)\citenamefont {Fu},
  \citenamefont {Torre}, \citenamefont {Willaime}, \citenamefont {Bocquet},\
  and\ \citenamefont {Barbu}}]{Fu2004}%
  \BibitemOpen
  \bibfield  {author} {\bibinfo {author} {\bibfnamefont {C.-C.}\ \bibnamefont
  {Fu}}, \bibinfo {author} {\bibfnamefont {J.~Dalla}\ \bibnamefont {Torre}},
  \bibinfo {author} {\bibfnamefont {F.}~\bibnamefont {Willaime}}, \bibinfo
  {author} {\bibfnamefont {J.-L.}\ \bibnamefont {Bocquet}}, \ and\ \bibinfo
  {author} {\bibfnamefont {A.}~\bibnamefont {Barbu}},\ }\bibfield  {title}
  {\enquote {\bibinfo {title} {Multiscale modelling of defect kinetics in
  irradiated iron},}\ }\href {\doibase 10.1038/nmat1286} {\bibfield  {journal}
  {\bibinfo  {journal} {Nat. Mater.}\ }\textbf {\bibinfo {volume} {4}},\
  \bibinfo {pages} {68--74} (\bibinfo {year} {2004})}\BibitemShut {NoStop}%
\bibitem [{\citenamefont {Fluss}\ \emph {et~al.}(2004)\citenamefont {Fluss},
  \citenamefont {Wirth}, \citenamefont {Wall}, \citenamefont {Felter},
  \citenamefont {Caturla}, \citenamefont {Kubota},\ and\ \citenamefont {de~la
  Rubia}}]{Fluss2004}%
  \BibitemOpen
  \bibfield  {author} {\bibinfo {author} {\bibfnamefont {M.~J.}\ \bibnamefont
  {Fluss}}, \bibinfo {author} {\bibfnamefont {B.~D.}\ \bibnamefont {Wirth}},
  \bibinfo {author} {\bibfnamefont {M.}~\bibnamefont {Wall}}, \bibinfo {author}
  {\bibfnamefont {T.~E.}\ \bibnamefont {Felter}}, \bibinfo {author}
  {\bibfnamefont {M.~J.}\ \bibnamefont {Caturla}}, \bibinfo {author}
  {\bibfnamefont {A.}~\bibnamefont {Kubota}}, \ and\ \bibinfo {author}
  {\bibfnamefont {T.~Diaz}\ \bibnamefont {de~la Rubia}},\ }\bibfield  {title}
  {\enquote {\bibinfo {title} {Temperature-dependent defect properties from
  ion-irradiation in {P}u({G}a)},}\ }\href {\doibase
  10.1016/j.jallcom.2003.08.080} {\bibfield  {journal} {\bibinfo  {journal} {J.
  Alloys Compd.}\ }\textbf {\bibinfo {volume} {368}},\ \bibinfo {pages}
  {62--74} (\bibinfo {year} {2004})}\BibitemShut {NoStop}%
\bibitem [{\citenamefont {Kinchin}\ and\ \citenamefont
  {Thompson}(1958)}]{Kinchin1958}%
  \BibitemOpen
  \bibfield  {author} {\bibinfo {author} {\bibfnamefont {G.~H.}\ \bibnamefont
  {Kinchin}}\ and\ \bibinfo {author} {\bibfnamefont {M.~W.}\ \bibnamefont
  {Thompson}},\ }\bibfield  {title} {\enquote {\bibinfo {title} {Irradiation
  damage and recovery in molybdenum and tungsten},}\ }\href {\doibase
  10.1016/0891-3919(58)90054-8} {\bibfield  {journal} {\bibinfo  {journal} {J.
  Nuclear Energy}\ }\textbf {\bibinfo {volume} {6}},\ \bibinfo {pages}
  {275--284} (\bibinfo {year} {1958})}\BibitemShut {NoStop}%
\bibitem [{\citenamefont {Isebeck}\ \emph {et~al.}(1966)\citenamefont
  {Isebeck}, \citenamefont {Rau}, \citenamefont {Schilling}, \citenamefont
  {Sonnenberg}, \citenamefont {Tischer},\ and\ \citenamefont
  {Wenzl}}]{Isebeck1966}%
  \BibitemOpen
  \bibfield  {author} {\bibinfo {author} {\bibfnamefont {K.}~\bibnamefont
  {Isebeck}}, \bibinfo {author} {\bibfnamefont {F.}~\bibnamefont {Rau}},
  \bibinfo {author} {\bibfnamefont {W.}~\bibnamefont {Schilling}}, \bibinfo
  {author} {\bibfnamefont {K.}~\bibnamefont {Sonnenberg}}, \bibinfo {author}
  {\bibfnamefont {P.}~\bibnamefont {Tischer}}, \ and\ \bibinfo {author}
  {\bibfnamefont {H.}~\bibnamefont {Wenzl}},\ }\bibfield  {title} {\enquote
  {\bibinfo {title} {Stored energy, volume, and resistivity change in neutron
  irradiated aluminium},}\ }\href@noop {} {\bibfield  {journal} {\bibinfo
  {journal} {Phys. Status Solidi B}\ }\textbf {\bibinfo {volume} {17}},\
  \bibinfo {pages} {259--268} (\bibinfo {year} {1966})}\BibitemShut {NoStop}%
\bibitem [{\citenamefont {Delaplace}\ \emph {et~al.}(1968)\citenamefont
  {Delaplace}, \citenamefont {Hillairet}, \citenamefont {Nicoud}, \citenamefont
  {Schumacher},\ and\ \citenamefont {Vogl}}]{Delaplace1968}%
  \BibitemOpen
  \bibfield  {author} {\bibinfo {author} {\bibfnamefont {J.}~\bibnamefont
  {Delaplace}}, \bibinfo {author} {\bibfnamefont {J.}~\bibnamefont
  {Hillairet}}, \bibinfo {author} {\bibfnamefont {J.~C.}\ \bibnamefont
  {Nicoud}}, \bibinfo {author} {\bibfnamefont {D.}~\bibnamefont {Schumacher}},
  \ and\ \bibinfo {author} {\bibfnamefont {G.}~\bibnamefont {Vogl}},\
  }\bibfield  {title} {\enquote {\bibinfo {title} {Low temperature neutron
  radiation damage and recovery in magnesium},}\ }\href {\doibase
  10.1002/pssb.19680300115} {\bibfield  {journal} {\bibinfo  {journal} {Phys.
  Status Solidi}\ }\textbf {\bibinfo {volume} {30}},\ \bibinfo {pages}
  {119--126} (\bibinfo {year} {1968})}\BibitemShut {NoStop}%
\bibitem [{\citenamefont {Nicoud}\ \emph {et~al.}(1968)\citenamefont {Nicoud},
  \citenamefont {Bonjour}, \citenamefont {Schumacher},\ and\ \citenamefont
  {Delaplace}}]{Nicoud1968}%
  \BibitemOpen
  \bibfield  {author} {\bibinfo {author} {\bibfnamefont {J.~C.}\ \bibnamefont
  {Nicoud}}, \bibinfo {author} {\bibfnamefont {E.}~\bibnamefont {Bonjour}},
  \bibinfo {author} {\bibfnamefont {D.}~\bibnamefont {Schumacher}}, \ and\
  \bibinfo {author} {\bibfnamefont {J.}~\bibnamefont {Delaplace}},\ }\bibfield
  {title} {\enquote {\bibinfo {title} {Restauration dans le stade i du
  beryllium irradie par des neutrons et par des electrons},}\ }\href {\doibase
  https://doi.org/10.1016/0375-9601(68)90616-6} {\bibfield  {journal} {\bibinfo
   {journal} {Phys. Lett. A}\ }\textbf {\bibinfo {volume} {26}},\ \bibinfo
  {pages} {228--229} (\bibinfo {year} {1968})}\BibitemShut {NoStop}%
\bibitem [{\citenamefont {Losehand}\ \emph {et~al.}(1969)\citenamefont
  {Losehand}, \citenamefont {Rau},\ and\ \citenamefont {Wenzl}}]{Losehand1969}%
  \BibitemOpen
  \bibfield  {author} {\bibinfo {author} {\bibfnamefont {R.}~\bibnamefont
  {Losehand}}, \bibinfo {author} {\bibfnamefont {F.}~\bibnamefont {Rau}}, \
  and\ \bibinfo {author} {\bibfnamefont {H.}~\bibnamefont {Wenzl}},\ }\bibfield
   {title} {\enquote {\bibinfo {title} {Stored energy and electrical
  resistivity of {F}renkel defects in copper},}\ }\href {\doibase
  10.1080/00337576908235586} {\bibfield  {journal} {\bibinfo  {journal}
  {Radiat. Eff.}\ }\textbf {\bibinfo {volume} {2}},\ \bibinfo {pages} {69--74}
  (\bibinfo {year} {1969})}\BibitemShut {NoStop}%
\bibitem [{\citenamefont {Takaki}\ \emph {et~al.}(1983)\citenamefont {Takaki},
  \citenamefont {Fuss}, \citenamefont {Kuglers}, \citenamefont {Dedek},\ and\
  \citenamefont {Schultz}}]{Takaki1983}%
  \BibitemOpen
  \bibfield  {author} {\bibinfo {author} {\bibfnamefont {S.}~\bibnamefont
  {Takaki}}, \bibinfo {author} {\bibfnamefont {J.}~\bibnamefont {Fuss}},
  \bibinfo {author} {\bibfnamefont {H.}~\bibnamefont {Kuglers}}, \bibinfo
  {author} {\bibfnamefont {U.}~\bibnamefont {Dedek}}, \ and\ \bibinfo {author}
  {\bibfnamefont {H.}~\bibnamefont {Schultz}},\ }\bibfield  {title} {\enquote
  {\bibinfo {title} {The resistivity recovery of high purity and carbon doped
  iron following low temperature electron irradiation},}\ }\href {\doibase
  10.1080/00337578308207398} {\bibfield  {journal} {\bibinfo  {journal}
  {Radiat. Eff. Defects Solids}\ }\textbf {\bibinfo {volume} {79}},\ \bibinfo
  {pages} {87--122} (\bibinfo {year} {1983})}\BibitemShut {NoStop}%
\bibitem [{\citenamefont {Gómez-Ferrer}(2016)}]{GomezFerrer2016}%
  \BibitemOpen
  \bibfield  {author} {\bibinfo {author} {\bibfnamefont {B.}~\bibnamefont
  {Gómez-Ferrer}},\ }\href {\doibase 10.1007/978-3-319-38857-1} {\emph
  {\bibinfo {title} {Resistivity Recovery in {Fe} and {FeCr} Alloys}}}\
  (\bibinfo  {publisher} {SpringerBriefs in Applied Sciences and Technology},\
  \bibinfo {year} {2016})\BibitemShut {NoStop}%
\bibitem [{\citenamefont {Nikolaev}(2007)}]{Nikolaev2007}%
  \BibitemOpen
  \bibfield  {author} {\bibinfo {author} {\bibfnamefont {A.~L.}\ \bibnamefont
  {Nikolaev}},\ }\bibfield  {title} {\enquote {\bibinfo {title} {Specificity of
  stage {III} in electron-irradiated {Fe}-{Cr} alloys},}\ }\href {\doibase
  10.1080/14786430701468977} {\bibfield  {journal} {\bibinfo  {journal} {Phil.
  Mag.}\ }\textbf {\bibinfo {volume} {87}},\ \bibinfo {pages} {4847--4874}
  (\bibinfo {year} {2007})}\BibitemShut {NoStop}%
\bibitem [{\citenamefont {Nikolaev}(2009)}]{Nikolaev2009}%
  \BibitemOpen
  \bibfield  {author} {\bibinfo {author} {\bibfnamefont {A.~L.}\ \bibnamefont
  {Nikolaev}},\ }\bibfield  {title} {\enquote {\bibinfo {title} {Difference
  approach to the analysis of resistivity recovery data for irradiated
  short-range ordered alloys},}\ }\href {\doibase 10.1080/14786430902835651}
  {\bibfield  {journal} {\bibinfo  {journal} {Phil. Mag.}\ }\textbf {\bibinfo
  {volume} {89}},\ \bibinfo {pages} {1017--1033} (\bibinfo {year}
  {2009})}\BibitemShut {NoStop}%
\bibitem [{\citenamefont {Nikolaev}(2018)}]{Nikolaev2018}%
  \BibitemOpen
  \bibfield  {author} {\bibinfo {author} {\bibfnamefont {A.~L.}\ \bibnamefont
  {Nikolaev}},\ }\bibfield  {title} {\enquote {\bibinfo {title} {The
  unambiguous identification of vacancy migration stage in resistivity recovery
  of irradiated metals based on trapping of vacancies at probe impurity
  atoms},}\ }\href {\doibase 10.1080/14786435.2018.1492158} {\bibfield
  {journal} {\bibinfo  {journal} {Phil. Mag.}\ }\textbf {\bibinfo {volume}
  {98}},\ \bibinfo {pages} {2481--2494} (\bibinfo {year} {2018})}\BibitemShut
  {NoStop}%
\bibitem [{\citenamefont {Bass}(1972)}]{Bass1972}%
  \BibitemOpen
  \bibfield  {author} {\bibinfo {author} {\bibfnamefont {J.}~\bibnamefont
  {Bass}},\ }\bibfield  {title} {\enquote {\bibinfo {title} {Deviations from
  {M}atthiessen's rule},}\ }\href {\doibase 10.1080/00018737200101308}
  {\bibfield  {journal} {\bibinfo  {journal} {Adv. Phys.}\ }\textbf {\bibinfo
  {volume} {21}},\ \bibinfo {pages} {431--604} (\bibinfo {year}
  {1972})}\BibitemShut {NoStop}%
\bibitem [{\citenamefont {Zinkle}(1988)}]{Zinkle1988}%
  \BibitemOpen
  \bibfield  {author} {\bibinfo {author} {\bibfnamefont {S.~J.}\ \bibnamefont
  {Zinkle}},\ }\bibfield  {title} {\enquote {\bibinfo {title} {Electrical
  resistivity of small dislocation loops in irradiated copper},}\ }\href
  {\doibase 10.1088/0305-4608/18/3/009} {\bibfield  {journal} {\bibinfo
  {journal} {J. Phys. F Met. Phys.}\ }\textbf {\bibinfo {volume} {18}},\
  \bibinfo {pages} {377--391} (\bibinfo {year} {1988})}\BibitemShut {NoStop}%
\bibitem [{\citenamefont {Mantl}\ and\ \citenamefont
  {Triftshäuser}(1978)}]{Mantl1978}%
  \BibitemOpen
  \bibfield  {author} {\bibinfo {author} {\bibfnamefont {S.}~\bibnamefont
  {Mantl}}\ and\ \bibinfo {author} {\bibfnamefont {W.}~\bibnamefont
  {Triftshäuser}},\ }\bibfield  {title} {\enquote {\bibinfo {title} {Defect
  annealing studies on metals by positron annihilation and electrical
  resitivity measurements},}\ }\href {\doibase 10.1103/physrevb.17.1645}
  {\bibfield  {journal} {\bibinfo  {journal} {Phys. Rev. B}\ }\textbf {\bibinfo
  {volume} {17}},\ \bibinfo {pages} {1645--1652} (\bibinfo {year}
  {1978})}\BibitemShut {NoStop}%
\bibitem [{\citenamefont {Eldrup}\ and\ \citenamefont
  {Singh}(1997)}]{Eldrup1997}%
  \BibitemOpen
  \bibfield  {author} {\bibinfo {author} {\bibfnamefont {M.}~\bibnamefont
  {Eldrup}}\ and\ \bibinfo {author} {\bibfnamefont {B.~N.}\ \bibnamefont
  {Singh}},\ }\bibfield  {title} {\enquote {\bibinfo {title} {Studies of
  defects and defect agglomerates by positron annihilation spectroscopy},}\
  }\href {\doibase 10.1016/s0022-3115(97)00221-3} {\bibfield  {journal}
  {\bibinfo  {journal} {J. Nucl. Mater.}\ }\textbf {\bibinfo {volume} {251}},\
  \bibinfo {pages} {132--138} (\bibinfo {year} {1997})}\BibitemShut {NoStop}%
\bibitem [{\citenamefont {Eldrup}\ and\ \citenamefont
  {Singh}(2003)}]{Eldrup2003}%
  \BibitemOpen
  \bibfield  {author} {\bibinfo {author} {\bibfnamefont {M.}~\bibnamefont
  {Eldrup}}\ and\ \bibinfo {author} {\bibfnamefont {B.~N.}\ \bibnamefont
  {Singh}},\ }\bibfield  {title} {\enquote {\bibinfo {title} {Accumulation of
  point defects and their complexes in irradiated metals as studied by the use
  of positron annihilation spectroscopy – a brief review},}\ }\href {\doibase
  10.1016/j.jnucmat.2003.08.011} {\bibfield  {journal} {\bibinfo  {journal} {J.
  Nucl. Mater.}\ }\textbf {\bibinfo {volume} {323}},\ \bibinfo {pages}
  {346--353} (\bibinfo {year} {2003})}\BibitemShut {NoStop}%
\bibitem [{\citenamefont {Hu}\ \emph {et~al.}(2016)\citenamefont {Hu},
  \citenamefont {Koyanagi}, \citenamefont {Fukuda}, \citenamefont {Katoh},
  \citenamefont {Snead},\ and\ \citenamefont {Wirth}}]{Hu2016}%
  \BibitemOpen
  \bibfield  {author} {\bibinfo {author} {\bibfnamefont {X.}~\bibnamefont
  {Hu}}, \bibinfo {author} {\bibfnamefont {T.}~\bibnamefont {Koyanagi}},
  \bibinfo {author} {\bibfnamefont {M.}~\bibnamefont {Fukuda}}, \bibinfo
  {author} {\bibfnamefont {Y.}~\bibnamefont {Katoh}}, \bibinfo {author}
  {\bibfnamefont {L.~L.}\ \bibnamefont {Snead}}, \ and\ \bibinfo {author}
  {\bibfnamefont {B.~D.}\ \bibnamefont {Wirth}},\ }\bibfield  {title} {\enquote
  {\bibinfo {title} {Defect evolution in single crystalline tungsten following
  low temperature and low dose neutron irradiation},}\ }\href {\doibase
  10.1016/j.jnucmat.2015.12.040} {\bibfield  {journal} {\bibinfo  {journal} {J.
  Nucl. Mater.}\ }\textbf {\bibinfo {volume} {470}},\ \bibinfo {pages}
  {278--289} (\bibinfo {year} {2016})}\BibitemShut {NoStop}%
\bibitem [{\citenamefont {Nagai}\ \emph {et~al.}(2003)\citenamefont {Nagai},
  \citenamefont {Takadate}, \citenamefont {Tang}, \citenamefont {Ohkubo},
  \citenamefont {Sunaga}, \citenamefont {Takizawa},\ and\ \citenamefont
  {Hasegawa}}]{Nagai2003}%
  \BibitemOpen
  \bibfield  {author} {\bibinfo {author} {\bibfnamefont {Y.}~\bibnamefont
  {Nagai}}, \bibinfo {author} {\bibfnamefont {K.}~\bibnamefont {Takadate}},
  \bibinfo {author} {\bibfnamefont {Z.}~\bibnamefont {Tang}}, \bibinfo {author}
  {\bibfnamefont {H.}~\bibnamefont {Ohkubo}}, \bibinfo {author} {\bibfnamefont
  {H.}~\bibnamefont {Sunaga}}, \bibinfo {author} {\bibfnamefont
  {H.}~\bibnamefont {Takizawa}}, \ and\ \bibinfo {author} {\bibfnamefont
  {M.}~\bibnamefont {Hasegawa}},\ }\bibfield  {title} {\enquote {\bibinfo
  {title} {Positron annihilation study of vacancy-solute complex evolution in
  {F}e-based alloys},}\ }\href@noop {} {\bibfield  {journal} {\bibinfo
  {journal} {Phys. Rev. B}\ }\textbf {\bibinfo {volume} {67}} (\bibinfo {year}
  {2003})}\BibitemShut {NoStop}%
\bibitem [{\citenamefont {Lynn}\ \emph {et~al.}(1986)\citenamefont {Lynn},
  \citenamefont {Chen}, \citenamefont {Nielsen}, \citenamefont {Pareja},\ and\
  \citenamefont {Myers}}]{Lynn1986}%
  \BibitemOpen
  \bibfield  {author} {\bibinfo {author} {\bibfnamefont {K.~G.}\ \bibnamefont
  {Lynn}}, \bibinfo {author} {\bibfnamefont {D.~M.}\ \bibnamefont {Chen}},
  \bibinfo {author} {\bibfnamefont {B.}~\bibnamefont {Nielsen}}, \bibinfo
  {author} {\bibfnamefont {R.}~\bibnamefont {Pareja}}, \ and\ \bibinfo {author}
  {\bibfnamefont {S.}~\bibnamefont {Myers}},\ }\bibfield  {title} {\enquote
  {\bibinfo {title} {Variable-energy positron-beam studies of {Ni} implanted
  with {He}},}\ }\href {\doibase 10.1103/physrevb.34.1449} {\bibfield
  {journal} {\bibinfo  {journal} {Phys. Rev. B}\ }\textbf {\bibinfo {volume}
  {34}},\ \bibinfo {pages} {1449--1458} (\bibinfo {year} {1986})}\BibitemShut
  {NoStop}%
\bibitem [{\citenamefont {Siemek}\ \emph {et~al.}(2021)\citenamefont {Siemek},
  \citenamefont {Horodek}, \citenamefont {Skuratov}, \citenamefont
  {Waliszewski},\ and\ \citenamefont {Sohatsky}}]{Siemek2021}%
  \BibitemOpen
  \bibfield  {author} {\bibinfo {author} {\bibfnamefont {K.}~\bibnamefont
  {Siemek}}, \bibinfo {author} {\bibfnamefont {P.}~\bibnamefont {Horodek}},
  \bibinfo {author} {\bibfnamefont {V.~A.}\ \bibnamefont {Skuratov}}, \bibinfo
  {author} {\bibfnamefont {J.}~\bibnamefont {Waliszewski}}, \ and\ \bibinfo
  {author} {\bibfnamefont {A.}~\bibnamefont {Sohatsky}},\ }\bibfield  {title}
  {\enquote {\bibinfo {title} {Positron annihilation studies of irradiation
  induced defects in nanostructured titanium},}\ }\href {\doibase
  10.1016/j.vacuum.2021.110282} {\bibfield  {journal} {\bibinfo  {journal}
  {Vacuum}\ }\textbf {\bibinfo {volume} {190}},\ \bibinfo {pages} {110282}
  (\bibinfo {year} {2021})}\BibitemShut {NoStop}%
\bibitem [{\citenamefont {Tuomisto}\ \emph {et~al.}(2020)\citenamefont
  {Tuomisto}, \citenamefont {Makkonen}, \citenamefont {Heikinheimo},
  \citenamefont {Granberg}, \citenamefont {Djurabekova}, \citenamefont
  {Nordlund}, \citenamefont {Velisa}, \citenamefont {Bei}, \citenamefont {Xue},
  \citenamefont {Weber},\ and\ \citenamefont {Zhang}}]{Tuomisto2020}%
  \BibitemOpen
  \bibfield  {author} {\bibinfo {author} {\bibfnamefont {F.}~\bibnamefont
  {Tuomisto}}, \bibinfo {author} {\bibfnamefont {I.}~\bibnamefont {Makkonen}},
  \bibinfo {author} {\bibfnamefont {J.}~\bibnamefont {Heikinheimo}}, \bibinfo
  {author} {\bibfnamefont {F.}~\bibnamefont {Granberg}}, \bibinfo {author}
  {\bibfnamefont {F.}~\bibnamefont {Djurabekova}}, \bibinfo {author}
  {\bibfnamefont {K.}~\bibnamefont {Nordlund}}, \bibinfo {author}
  {\bibfnamefont {G.}~\bibnamefont {Velisa}}, \bibinfo {author} {\bibfnamefont
  {H.}~\bibnamefont {Bei}}, \bibinfo {author} {\bibfnamefont {H.}~\bibnamefont
  {Xue}}, \bibinfo {author} {\bibfnamefont {W.~J.}\ \bibnamefont {Weber}}, \
  and\ \bibinfo {author} {\bibfnamefont {Y.}~\bibnamefont {Zhang}},\ }\bibfield
   {title} {\enquote {\bibinfo {title} {Segregation of ni at early stages of
  radiation damage in {NiCoFeCr} solid solution alloys},}\ }\href {\doibase
  10.1016/j.actamat.2020.06.024} {\bibfield  {journal} {\bibinfo  {journal}
  {Acta Mater.}\ }\textbf {\bibinfo {volume} {196}},\ \bibinfo {pages} {44--51}
  (\bibinfo {year} {2020})}\BibitemShut {NoStop}%
\bibitem [{\citenamefont {Soneda}\ \emph {et~al.}(2003)\citenamefont {Soneda},
  \citenamefont {Ishino}, \citenamefont {Takahashi},\ and\ \citenamefont
  {Dohi}}]{Soneda2003}%
  \BibitemOpen
  \bibfield  {author} {\bibinfo {author} {\bibfnamefont {N.}~\bibnamefont
  {Soneda}}, \bibinfo {author} {\bibfnamefont {S.}~\bibnamefont {Ishino}},
  \bibinfo {author} {\bibfnamefont {A.}~\bibnamefont {Takahashi}}, \ and\
  \bibinfo {author} {\bibfnamefont {K.}~\bibnamefont {Dohi}},\ }\bibfield
  {title} {\enquote {\bibinfo {title} {Modeling the microstructural evolution
  in bcc-{Fe} during irradiation using kinetic {M}onte {C}arlo computer
  simulation},}\ }\href {\doibase 10.1016/j.jnucmat.2003.08.021} {\bibfield
  {journal} {\bibinfo  {journal} {J. Nucl. Mater.}\ }\textbf {\bibinfo {volume}
  {323}},\ \bibinfo {pages} {169--180} (\bibinfo {year} {2003})}\BibitemShut
  {NoStop}%
\bibitem [{\citenamefont {Andrianov}\ \emph {et~al.}(2021)\citenamefont
  {Andrianov}, \citenamefont {Bedelbekova},\ and\ \citenamefont
  {Trigub}}]{Andrianov2021}%
  \BibitemOpen
  \bibfield  {author} {\bibinfo {author} {\bibfnamefont {V.~A.}\ \bibnamefont
  {Andrianov}}, \bibinfo {author} {\bibfnamefont {K.~A.}\ \bibnamefont
  {Bedelbekova}}, \ and\ \bibinfo {author} {\bibfnamefont {A.~L.}\ \bibnamefont
  {Trigub}},\ }\bibfield  {title} {\enquote {\bibinfo {title} {Study of
  radiation defects in metal {Mo} and {Ta} by {M}össbauer effect and
  {EXAFS}},}\ }\href {\doibase 10.1016/j.vacuum.2021.110521} {\bibfield
  {journal} {\bibinfo  {journal} {Vacuum}\ }\textbf {\bibinfo {volume} {193}},\
  \bibinfo {pages} {110521} (\bibinfo {year} {2021})}\BibitemShut {NoStop}%
\bibitem [{\citenamefont {Mason}\ \emph {et~al.}(2021)\citenamefont {Mason},
  \citenamefont {Reza}, \citenamefont {Granberg},\ and\ \citenamefont
  {Hofmann}}]{Mason2021}%
  \BibitemOpen
  \bibfield  {author} {\bibinfo {author} {\bibfnamefont {D.~R.}\ \bibnamefont
  {Mason}}, \bibinfo {author} {\bibfnamefont {A.}~\bibnamefont {Reza}},
  \bibinfo {author} {\bibfnamefont {F.}~\bibnamefont {Granberg}}, \ and\
  \bibinfo {author} {\bibfnamefont {F.}~\bibnamefont {Hofmann}},\ }\bibfield
  {title} {\enquote {\bibinfo {title} {Estimate for thermal diffusivity in
  highly irradiated tungsten using molecular dynamics simulation},}\ }\href
  {\doibase 10.1103/physrevmaterials.5.125407} {\bibfield  {journal} {\bibinfo
  {journal} {Phys. Rev. Mater.}\ }\textbf {\bibinfo {volume} {5}},\ \bibinfo
  {pages} {125407} (\bibinfo {year} {2021})}\BibitemShut {NoStop}%
\bibitem [{\citenamefont {Larson}(1975)}]{Larson1975}%
  \BibitemOpen
  \bibfield  {author} {\bibinfo {author} {\bibfnamefont {B.~C.}\ \bibnamefont
  {Larson}},\ }\bibfield  {title} {\enquote {\bibinfo {title} {{X}-ray studies
  of defect clusters in copper},}\ }\href {\doibase 10.1107/s0021889875009946}
  {\bibfield  {journal} {\bibinfo  {journal} {J. Appl. Cryst.}\ }\textbf
  {\bibinfo {volume} {8}},\ \bibinfo {pages} {150--160} (\bibinfo {year}
  {1975})}\BibitemShut {NoStop}%
\bibitem [{\citenamefont {Larson}\ and\ \citenamefont
  {Young}(1987)}]{Larson1987}%
  \BibitemOpen
  \bibfield  {author} {\bibinfo {author} {\bibfnamefont {B.~C.}\ \bibnamefont
  {Larson}}\ and\ \bibinfo {author} {\bibfnamefont {F.~W.}\ \bibnamefont
  {Young}},\ }\bibfield  {title} {\enquote {\bibinfo {title} {{X}-ray diffuse
  scattering study of irradiation induced dislocation loops in copper},}\
  }\href {\doibase 10.1002/pssa.2211040120} {\bibfield  {journal} {\bibinfo
  {journal} {Phys. Status Solidi A}\ }\textbf {\bibinfo {volume} {104}},\
  \bibinfo {pages} {273--286} (\bibinfo {year} {1987})}\BibitemShut {NoStop}%
\bibitem [{\citenamefont {Ehrhart}\ and\ \citenamefont
  {Averback}(1989)}]{Ehrhart1989}%
  \BibitemOpen
  \bibfield  {author} {\bibinfo {author} {\bibfnamefont {P.}~\bibnamefont
  {Ehrhart}}\ and\ \bibinfo {author} {\bibfnamefont {R.~S.}\ \bibnamefont
  {Averback}},\ }\bibfield  {title} {\enquote {\bibinfo {title} {Diffuse
  {X}-ray scattering studies of neutron- and electron-irradiated {Ni}, {Cu} and
  dilute alloys},}\ }\href {\doibase 10.1080/01418618908213863} {\bibfield
  {journal} {\bibinfo  {journal} {Phil. Mag.}\ }\textbf {\bibinfo {volume}
  {60}},\ \bibinfo {pages} {283--306} (\bibinfo {year} {1989})}\BibitemShut
  {NoStop}%
\bibitem [{\citenamefont {Narayan}\ and\ \citenamefont
  {Larson}(1977)}]{Narayan1977}%
  \BibitemOpen
  \bibfield  {author} {\bibinfo {author} {\bibfnamefont {J.}~\bibnamefont
  {Narayan}}\ and\ \bibinfo {author} {\bibfnamefont {B.~C.}\ \bibnamefont
  {Larson}},\ }\bibfield  {title} {\enquote {\bibinfo {title} {Defect clusters
  and annealing in self‐ion‐irradiated nickel},}\ }\href {\doibase
  10.1063/1.323468} {\bibfield  {journal} {\bibinfo  {journal} {J. Appl.
  Phys.}\ }\textbf {\bibinfo {volume} {48}},\ \bibinfo {pages} {4536--4539}
  (\bibinfo {year} {1977})}\BibitemShut {NoStop}%
\bibitem [{\citenamefont {H.Yuya}\ \emph {et~al.}(1999)\citenamefont {H.Yuya},
  \citenamefont {Maeta}, \citenamefont {Ohtsuka}, \citenamefont {Matsumoto},
  \citenamefont {Sugai}, \citenamefont {Iwase}, \citenamefont {Matsui},
  \citenamefont {Suzuki}, \citenamefont {Jinchoh},\ and\ \citenamefont
  {Yamakawa}}]{Yuya1999}%
  \BibitemOpen
  \bibfield  {author} {\bibinfo {author} {\bibnamefont {H.Yuya}}, \bibinfo
  {author} {\bibfnamefont {H.}~\bibnamefont {Maeta}}, \bibinfo {author}
  {\bibfnamefont {H.}~\bibnamefont {Ohtsuka}}, \bibinfo {author} {\bibfnamefont
  {N.}~\bibnamefont {Matsumoto}}, \bibinfo {author} {\bibfnamefont
  {H.}~\bibnamefont {Sugai}}, \bibinfo {author} {\bibfnamefont
  {A.}~\bibnamefont {Iwase}}, \bibinfo {author} {\bibfnamefont
  {T.}~\bibnamefont {Matsui}}, \bibinfo {author} {\bibfnamefont
  {T.}~\bibnamefont {Suzuki}}, \bibinfo {author} {\bibfnamefont
  {M.}~\bibnamefont {Jinchoh}}, \ and\ \bibinfo {author} {\bibfnamefont
  {K.}~\bibnamefont {Yamakawa}},\ }\bibfield  {title} {\enquote {\bibinfo
  {title} {Diffuse {X}-ray scattering studies of radiation defects in {Ni} and
  dilute {Ni} alloys},}\ }\href {\doibase 10.1016/s0022-3115(98)00642-4}
  {\bibfield  {journal} {\bibinfo  {journal} {Nucl. Instrum. Methods. Phys.
  Res. B}\ }\textbf {\bibinfo {volume} {271-272}},\ \bibinfo {pages} {7--10}
  (\bibinfo {year} {1999})}\BibitemShut {NoStop}%
\bibitem [{\citenamefont {Sun}\ \emph {et~al.}(2018)\citenamefont {Sun},
  \citenamefont {Wang}, \citenamefont {Frost}, \citenamefont {Schönwälder},
  \citenamefont {Levitan}, \citenamefont {Mo}, \citenamefont {Chen},
  \citenamefont {Hastings}, \citenamefont {Tynan}, \citenamefont {Glenzer},\
  and\ \citenamefont {Heimann}}]{Sun2018}%
  \BibitemOpen
  \bibfield  {author} {\bibinfo {author} {\bibfnamefont {P.}~\bibnamefont
  {Sun}}, \bibinfo {author} {\bibfnamefont {Y.}~\bibnamefont {Wang}}, \bibinfo
  {author} {\bibfnamefont {M.}~\bibnamefont {Frost}}, \bibinfo {author}
  {\bibfnamefont {C.}~\bibnamefont {Schönwälder}}, \bibinfo {author}
  {\bibfnamefont {A.~L.}\ \bibnamefont {Levitan}}, \bibinfo {author}
  {\bibfnamefont {M.}~\bibnamefont {Mo}}, \bibinfo {author} {\bibfnamefont
  {Z.}~\bibnamefont {Chen}}, \bibinfo {author} {\bibfnamefont {J.~B.}\
  \bibnamefont {Hastings}}, \bibinfo {author} {\bibfnamefont {G.~R.}\
  \bibnamefont {Tynan}}, \bibinfo {author} {\bibfnamefont {S.~H.}\ \bibnamefont
  {Glenzer}}, \ and\ \bibinfo {author} {\bibfnamefont {P.}~\bibnamefont
  {Heimann}},\ }\bibfield  {title} {\enquote {\bibinfo {title}
  {Characterization of defect clusters in ion-irradiated tungsten by {X}-ray
  diffuse scattering},}\ }\href {\doibase 10.1016/j.jnucmat.2018.07.062}
  {\bibfield  {journal} {\bibinfo  {journal} {J. Nucl. Mater.}\ }\textbf
  {\bibinfo {volume} {510}},\ \bibinfo {pages} {322--330} (\bibinfo {year}
  {2018})}\BibitemShut {NoStop}%
\bibitem [{\citenamefont {Ribárik}(2008)}]{Ribarik2008}%
  \BibitemOpen
  \bibfield  {author} {\bibinfo {author} {\bibfnamefont {G.}~\bibnamefont
  {Ribárik}},\ }\emph {\bibinfo {title} {Modeling of diffraction patterns
  based on microstructural properties}},\ \href@noop {} {Ph.D. thesis},\
  \bibinfo  {school} {Eötvös Loránd University} (\bibinfo {year}
  {2008})\BibitemShut {NoStop}%
\bibitem [{\citenamefont {Ribárik}\ \emph {et~al.}(2020)\citenamefont
  {Ribárik}, \citenamefont {Jóni},\ and\ \citenamefont
  {Ungár}}]{Ribarik2020}%
  \BibitemOpen
  \bibfield  {author} {\bibinfo {author} {\bibfnamefont {G.}~\bibnamefont
  {Ribárik}}, \bibinfo {author} {\bibfnamefont {B.}~\bibnamefont {Jóni}}, \
  and\ \bibinfo {author} {\bibfnamefont {T.}~\bibnamefont {Ungár}},\
  }\bibfield  {title} {\enquote {\bibinfo {title} {The convolutional multiple
  whole profile {(CMWP)} fitting method, a global optimization procedure for
  microstructure determination},}\ }\href {\doibase 10.3390/cryst10070623}
  {\bibfield  {journal} {\bibinfo  {journal} {Crystals}\ }\textbf {\bibinfo
  {volume} {10}},\ \bibinfo {pages} {623} (\bibinfo {year} {2020})}\BibitemShut
  {NoStop}%
\bibitem [{\citenamefont {Seymour}\ \emph {et~al.}(2017)\citenamefont
  {Seymour}, \citenamefont {Frankel}, \citenamefont {Balogh}, \citenamefont
  {Ungár}, \citenamefont {Thompson}, \citenamefont {Jädernäs}, \citenamefont
  {Romero}, \citenamefont {Hallstadius}, \citenamefont {Daymond}, \citenamefont
  {Ribárik},\ and\ \citenamefont {Preuss}}]{Seymour2017}%
  \BibitemOpen
  \bibfield  {author} {\bibinfo {author} {\bibfnamefont {T.}~\bibnamefont
  {Seymour}}, \bibinfo {author} {\bibfnamefont {P.}~\bibnamefont {Frankel}},
  \bibinfo {author} {\bibfnamefont {L.}~\bibnamefont {Balogh}}, \bibinfo
  {author} {\bibfnamefont {T.}~\bibnamefont {Ungár}}, \bibinfo {author}
  {\bibfnamefont {S.~P.}\ \bibnamefont {Thompson}}, \bibinfo {author}
  {\bibfnamefont {D.}~\bibnamefont {Jädernäs}}, \bibinfo {author}
  {\bibfnamefont {J.}~\bibnamefont {Romero}}, \bibinfo {author} {\bibfnamefont
  {L.}~\bibnamefont {Hallstadius}}, \bibinfo {author} {\bibfnamefont {M.~R.}\
  \bibnamefont {Daymond}}, \bibinfo {author} {\bibfnamefont {G.}~\bibnamefont
  {Ribárik}}, \ and\ \bibinfo {author} {\bibfnamefont {M.}~\bibnamefont
  {Preuss}},\ }\bibfield  {title} {\enquote {\bibinfo {title} {Evolution of
  dislocation structure in neutron irradiated {Z}ircaloy-2 studied by
  synchrotron {X}-ray diffraction peak profile analysis},}\ }\href {\doibase
  10.1016/j.actamat.2016.12.031} {\bibfield  {journal} {\bibinfo  {journal}
  {Acta Mater.}\ }\textbf {\bibinfo {volume} {126}},\ \bibinfo {pages}
  {102--113} (\bibinfo {year} {2017})}\BibitemShut {NoStop}%
\bibitem [{\citenamefont {Ungár}\ \emph
  {et~al.}(2021{\natexlab{b}})\citenamefont {Ungár}, \citenamefont {Ribárik},
  \citenamefont {Topping}, \citenamefont {Jones}, \citenamefont {Xu},
  \citenamefont {Hulse}, \citenamefont {Harte}, \citenamefont {Tichy},
  \citenamefont {Race}, \citenamefont {Frankel},\ and\ \citenamefont
  {Preuss}}]{Ungar2021a}%
  \BibitemOpen
  \bibfield  {author} {\bibinfo {author} {\bibfnamefont {T.}~\bibnamefont
  {Ungár}}, \bibinfo {author} {\bibfnamefont {G.}~\bibnamefont {Ribárik}},
  \bibinfo {author} {\bibfnamefont {M.}~\bibnamefont {Topping}}, \bibinfo
  {author} {\bibfnamefont {R.~M.~A.}\ \bibnamefont {Jones}}, \bibinfo {author}
  {\bibfnamefont {X.~D.}\ \bibnamefont {Xu}}, \bibinfo {author} {\bibfnamefont
  {R.}~\bibnamefont {Hulse}}, \bibinfo {author} {\bibfnamefont
  {A.}~\bibnamefont {Harte}}, \bibinfo {author} {\bibfnamefont
  {G.}~\bibnamefont {Tichy}}, \bibinfo {author} {\bibfnamefont {C.~P.}\
  \bibnamefont {Race}}, \bibinfo {author} {\bibfnamefont {P.}~\bibnamefont
  {Frankel}}, \ and\ \bibinfo {author} {\bibfnamefont {M.}~\bibnamefont
  {Preuss}},\ }\bibfield  {title} {\enquote {\bibinfo {title} {Characterizing
  dislocation loops in irradiated polycrystalline {Zr} alloys by {X}-ray line
  profile analysis of powder diffraction patterns with satellites},}\ }\href
  {\doibase 10.1107/s1600576721002673} {\bibfield  {journal} {\bibinfo
  {journal} {J. Appl. Cryst.}\ }\textbf {\bibinfo {volume} {54}},\ \bibinfo
  {pages} {803--821} (\bibinfo {year} {2021}{\natexlab{b}})}\BibitemShut
  {NoStop}%
\bibitem [{\citenamefont {Booth}\ \emph {et~al.}(2007)\citenamefont {Booth},
  \citenamefont {Bauer}, \citenamefont {Daniel}, \citenamefont {Wilson},
  \citenamefont {Mitchell}, \citenamefont {Morales}, \citenamefont {Sarrao},\
  and\ \citenamefont {Allen}}]{Booth2007}%
  \BibitemOpen
  \bibfield  {author} {\bibinfo {author} {\bibfnamefont {C.~H.}\ \bibnamefont
  {Booth}}, \bibinfo {author} {\bibfnamefont {E.~D.}\ \bibnamefont {Bauer}},
  \bibinfo {author} {\bibfnamefont {M.}~\bibnamefont {Daniel}}, \bibinfo
  {author} {\bibfnamefont {R.~E.}\ \bibnamefont {Wilson}}, \bibinfo {author}
  {\bibfnamefont {J.~N.}\ \bibnamefont {Mitchell}}, \bibinfo {author}
  {\bibfnamefont {L.~A.}\ \bibnamefont {Morales}}, \bibinfo {author}
  {\bibfnamefont {J.~L.}\ \bibnamefont {Sarrao}}, \ and\ \bibinfo {author}
  {\bibfnamefont {P.~G.}\ \bibnamefont {Allen}},\ }\bibfield  {title} {\enquote
  {\bibinfo {title} {Quantifying structural damage from self-irradiation in a
  plutonium superconductor},}\ }\href@noop {} {\bibfield  {journal} {\bibinfo
  {journal} {Phys. Rev. B}\ }\textbf {\bibinfo {volume} {76}} (\bibinfo {year}
  {2007})}\BibitemShut {NoStop}%
\bibitem [{\citenamefont {Booth}\ \emph {et~al.}(2013)\citenamefont {Booth},
  \citenamefont {Jiang}, \citenamefont {Medling}, \citenamefont {Wang},
  \citenamefont {Costello}, \citenamefont {Schwartz}, \citenamefont {Mitchell},
  \citenamefont {Tobash}, \citenamefont {Bauer}, \citenamefont {McCall},
  \citenamefont {Wall},\ and\ \citenamefont {Allen}}]{Booth2013}%
  \BibitemOpen
  \bibfield  {author} {\bibinfo {author} {\bibfnamefont {C.~H.}\ \bibnamefont
  {Booth}}, \bibinfo {author} {\bibfnamefont {Yu}~\bibnamefont {Jiang}},
  \bibinfo {author} {\bibfnamefont {S.~A.}\ \bibnamefont {Medling}}, \bibinfo
  {author} {\bibfnamefont {D.~L.}\ \bibnamefont {Wang}}, \bibinfo {author}
  {\bibfnamefont {A.~L.}\ \bibnamefont {Costello}}, \bibinfo {author}
  {\bibfnamefont {D.~S.}\ \bibnamefont {Schwartz}}, \bibinfo {author}
  {\bibfnamefont {J.~N.}\ \bibnamefont {Mitchell}}, \bibinfo {author}
  {\bibfnamefont {P.~H.}\ \bibnamefont {Tobash}}, \bibinfo {author}
  {\bibfnamefont {E.~D.}\ \bibnamefont {Bauer}}, \bibinfo {author}
  {\bibfnamefont {S.~K.}\ \bibnamefont {McCall}}, \bibinfo {author}
  {\bibfnamefont {M.~A.}\ \bibnamefont {Wall}}, \ and\ \bibinfo {author}
  {\bibfnamefont {P.~G.}\ \bibnamefont {Allen}},\ }\bibfield  {title} {\enquote
  {\bibinfo {title} {Self-irradiation damage to the local structure of
  plutonium and plutonium intermetallics},}\ }\href {\doibase
  10.1063/1.4794016} {\bibfield  {journal} {\bibinfo  {journal} {J. Appl.
  Phys.}\ }\textbf {\bibinfo {volume} {113}},\ \bibinfo {pages} {093502}
  (\bibinfo {year} {2013})}\BibitemShut {NoStop}%
\bibitem [{\citenamefont {Olive}\ \emph {et~al.}(2016)\citenamefont {Olive},
  \citenamefont {Wang}, \citenamefont {Booth}, \citenamefont {Bauer},
  \citenamefont {Pugmire}, \citenamefont {Freibert}, \citenamefont {McCall},
  \citenamefont {Wall},\ and\ \citenamefont {Allen}}]{Olive2016}%
  \BibitemOpen
  \bibfield  {author} {\bibinfo {author} {\bibfnamefont {D.~T.}\ \bibnamefont
  {Olive}}, \bibinfo {author} {\bibfnamefont {D.~L.}\ \bibnamefont {Wang}},
  \bibinfo {author} {\bibfnamefont {C.~H.}\ \bibnamefont {Booth}}, \bibinfo
  {author} {\bibfnamefont {E.~D.}\ \bibnamefont {Bauer}}, \bibinfo {author}
  {\bibfnamefont {A.~L.}\ \bibnamefont {Pugmire}}, \bibinfo {author}
  {\bibfnamefont {F.~J.}\ \bibnamefont {Freibert}}, \bibinfo {author}
  {\bibfnamefont {S.~K.}\ \bibnamefont {McCall}}, \bibinfo {author}
  {\bibfnamefont {M.~A.}\ \bibnamefont {Wall}}, \ and\ \bibinfo {author}
  {\bibfnamefont {P.~G.}\ \bibnamefont {Allen}},\ }\bibfield  {title} {\enquote
  {\bibinfo {title} {Isochronal annealing effects on local structure,
  crystalline fraction, and undamaged region size of radiation damage in
  {Ga}-stabilized delta-{Pu}},}\ }\href {\doibase 10.1063/1.4958856} {\bibfield
   {journal} {\bibinfo  {journal} {J. Appl. Phys.}\ }\textbf {\bibinfo {volume}
  {120}},\ \bibinfo {pages} {035103} (\bibinfo {year} {2016})}\BibitemShut
  {NoStop}%
\bibitem [{\citenamefont {Booth}\ and\ \citenamefont
  {Olive}(2017)}]{Booth2017}%
  \BibitemOpen
  \bibfield  {author} {\bibinfo {author} {\bibfnamefont {C.~H.}\ \bibnamefont
  {Booth}}\ and\ \bibinfo {author} {\bibfnamefont {D.~T.}\ \bibnamefont
  {Olive}},\ }\bibfield  {title} {\enquote {\bibinfo {title} {Effect of
  temperature and radiation damage on the local atomic structure of metallic
  plutonium and related compounds},}\ }\href {\doibase
  10.1080/23746149.2016.1243994} {\bibfield  {journal} {\bibinfo  {journal}
  {Adv. Phys. X}\ }\textbf {\bibinfo {volume} {2}},\ \bibinfo {pages} {1--21}
  (\bibinfo {year} {2017})}\BibitemShut {NoStop}%
\bibitem [{\citenamefont {Okamoto}\ \emph {et~al.}(1991)\citenamefont
  {Okamoto}, \citenamefont {Takagi},\ and\ \citenamefont
  {Kawamura}}]{Okamoto1991}%
  \BibitemOpen
  \bibfield  {author} {\bibinfo {author} {\bibfnamefont {Y.}~\bibnamefont
  {Okamoto}}, \bibinfo {author} {\bibfnamefont {R.}~\bibnamefont {Takagi}}, \
  and\ \bibinfo {author} {\bibfnamefont {K.}~\bibnamefont {Kawamura}},\
  }\bibfield  {title} {\enquote {\bibinfo {title} {Molecular dynamics of
  radiation damage in amorphous {Pd$_{80}$Si$_{20}$} alloy using {N$^+$}
  ions},}\ }\href {\doibase 10.1080/08927029108022443} {\bibfield  {journal}
  {\bibinfo  {journal} {Molecular Simulation}\ }\textbf {\bibinfo {volume}
  {6}},\ \bibinfo {pages} {353--362} (\bibinfo {year} {1991})}\BibitemShut
  {NoStop}%
\bibitem [{\citenamefont {Kuri}\ \emph {et~al.}(2009)\citenamefont {Kuri},
  \citenamefont {Cammelli}, \citenamefont {Degueldre}, \citenamefont
  {Bertsch},\ and\ \citenamefont {Gavillet}}]{Kuri2009}%
  \BibitemOpen
  \bibfield  {author} {\bibinfo {author} {\bibfnamefont {G.}~\bibnamefont
  {Kuri}}, \bibinfo {author} {\bibfnamefont {S.}~\bibnamefont {Cammelli}},
  \bibinfo {author} {\bibfnamefont {C.}~\bibnamefont {Degueldre}}, \bibinfo
  {author} {\bibfnamefont {J.}~\bibnamefont {Bertsch}}, \ and\ \bibinfo
  {author} {\bibfnamefont {D.}~\bibnamefont {Gavillet}},\ }\bibfield  {title}
  {\enquote {\bibinfo {title} {Neutron induced damage in reactor pressure
  vessel steel: {A}n {X}-ray absorption fine structure study},}\ }\href
  {\doibase 10.1016/j.jnucmat.2008.12.037} {\bibfield  {journal} {\bibinfo
  {journal} {J. Nucl. Mater.}\ }\textbf {\bibinfo {volume} {385}},\ \bibinfo
  {pages} {312--318} (\bibinfo {year} {2009})}\BibitemShut {NoStop}%
\bibitem [{\citenamefont {Chu}\ \emph {et~al.}(1973)\citenamefont {Chu},
  \citenamefont {Mayer}, \citenamefont {Nicolet}, \citenamefont {Buck},
  \citenamefont {Amsel},\ and\ \citenamefont {Eisen}}]{Chu1973}%
  \BibitemOpen
  \bibfield  {author} {\bibinfo {author} {\bibfnamefont {W.~K.}\ \bibnamefont
  {Chu}}, \bibinfo {author} {\bibfnamefont {J.~W.}\ \bibnamefont {Mayer}},
  \bibinfo {author} {\bibfnamefont {M.-A.}\ \bibnamefont {Nicolet}}, \bibinfo
  {author} {\bibfnamefont {T.~M.}\ \bibnamefont {Buck}}, \bibinfo {author}
  {\bibfnamefont {G.}~\bibnamefont {Amsel}}, \ and\ \bibinfo {author}
  {\bibfnamefont {F.}~\bibnamefont {Eisen}},\ }\bibfield  {title} {\enquote
  {\bibinfo {title} {Principles and applications of ion beam techniques for the
  analysis of solids and thin films},}\ }\href {\doibase
  10.1016/0040-6090(73)90002-3} {\bibfield  {journal} {\bibinfo  {journal}
  {Thin Solid Films}\ }\textbf {\bibinfo {volume} {17}},\ \bibinfo {pages}
  {1--41} (\bibinfo {year} {1973})}\BibitemShut {NoStop}%
\bibitem [{\citenamefont {Matzke}(1985)}]{Matzke1985}%
  \BibitemOpen
  \bibfield  {author} {\bibinfo {author} {\bibfnamefont {H.}~\bibnamefont
  {Matzke}},\ }\bibfield  {title} {\enquote {\bibinfo {title} {Application of
  ion beam techniques to solid state physics and technology of nuclear
  materials},}\ }\href {\doibase 10.1016/0022-3115(85)90001-7} {\bibfield
  {journal} {\bibinfo  {journal} {J. Nucl. Mater.}\ }\textbf {\bibinfo {volume}
  {136}},\ \bibinfo {pages} {143--153} (\bibinfo {year} {1985})}\BibitemShut
  {NoStop}%
\bibitem [{\citenamefont {Lu}\ \emph {et~al.}(2016)\citenamefont {Lu},
  \citenamefont {Jin}, \citenamefont {Béland}, \citenamefont {Zhang},
  \citenamefont {Yang}, \citenamefont {Qiao}, \citenamefont {Zhang},
  \citenamefont {Bei}, \citenamefont {Christen}, \citenamefont {Stoller},\ and\
  \citenamefont {Wang}}]{Lu2016}%
  \BibitemOpen
  \bibfield  {author} {\bibinfo {author} {\bibfnamefont {C.}~\bibnamefont
  {Lu}}, \bibinfo {author} {\bibfnamefont {K.}~\bibnamefont {Jin}}, \bibinfo
  {author} {\bibfnamefont {L.~K.}\ \bibnamefont {Béland}}, \bibinfo {author}
  {\bibfnamefont {F.}~\bibnamefont {Zhang}}, \bibinfo {author} {\bibfnamefont
  {T.}~\bibnamefont {Yang}}, \bibinfo {author} {\bibfnamefont {L.}~\bibnamefont
  {Qiao}}, \bibinfo {author} {\bibfnamefont {Y.}~\bibnamefont {Zhang}},
  \bibinfo {author} {\bibfnamefont {H.}~\bibnamefont {Bei}}, \bibinfo {author}
  {\bibfnamefont {H.~M.}\ \bibnamefont {Christen}}, \bibinfo {author}
  {\bibfnamefont {R.~E.}\ \bibnamefont {Stoller}}, \ and\ \bibinfo {author}
  {\bibfnamefont {L.}~\bibnamefont {Wang}},\ }\bibfield  {title} {\enquote
  {\bibinfo {title} {Direct observation of defect range and evolution in
  ion-irradiated single crystalline ni and ni binary alloys},}\ }\href
  {\doibase 10.1038/srep19994} {\bibfield  {journal} {\bibinfo  {journal} {Sci.
  Rep.}\ }\textbf {\bibinfo {volume} {6}} (\bibinfo {year} {2016}),\
  10.1038/srep19994}\BibitemShut {NoStop}%
\bibitem [{\citenamefont {Zhang}\ \emph {et~al.}(2015)\citenamefont {Zhang},
  \citenamefont {Stocks}, \citenamefont {Jin}, \citenamefont {Lu},
  \citenamefont {Bei}, \citenamefont {Sales}, \citenamefont {Wang},
  \citenamefont {B\'{e}land}, \citenamefont {Stoller}, \citenamefont
  {Samolyuk}, \citenamefont {Caro}, \citenamefont {Caro},\ and\ \citenamefont
  {Weber}}]{Zhang2015}%
  \BibitemOpen
  \bibfield  {author} {\bibinfo {author} {\bibfnamefont {Y.}~\bibnamefont
  {Zhang}}, \bibinfo {author} {\bibfnamefont {G.~M.}\ \bibnamefont {Stocks}},
  \bibinfo {author} {\bibfnamefont {K.}~\bibnamefont {Jin}}, \bibinfo {author}
  {\bibfnamefont {C.}~\bibnamefont {Lu}}, \bibinfo {author} {\bibfnamefont
  {H.}~\bibnamefont {Bei}}, \bibinfo {author} {\bibfnamefont {B.~C.}\
  \bibnamefont {Sales}}, \bibinfo {author} {\bibfnamefont {L.}~\bibnamefont
  {Wang}}, \bibinfo {author} {\bibfnamefont {L.~K.}\ \bibnamefont
  {B\'{e}land}}, \bibinfo {author} {\bibfnamefont {R.~E.}\ \bibnamefont
  {Stoller}}, \bibinfo {author} {\bibfnamefont {G.~D.}\ \bibnamefont
  {Samolyuk}}, \bibinfo {author} {\bibfnamefont {M.}~\bibnamefont {Caro}},
  \bibinfo {author} {\bibfnamefont {A.}~\bibnamefont {Caro}}, \ and\ \bibinfo
  {author} {\bibfnamefont {W.~J.}\ \bibnamefont {Weber}},\ }\bibfield  {title}
  {\enquote {\bibinfo {title} {Influence of chemical disorder on energy
  dissipation and defect evolution in concentrated solid solution alloys},}\
  }\href {\doibase 10.1038/ncomms9736} {\bibfield  {journal} {\bibinfo
  {journal} {Nat. Commun.}\ }\textbf {\bibinfo {volume} {6}},\ \bibinfo {pages}
  {8736} (\bibinfo {year} {2015})}\BibitemShut {NoStop}%
\bibitem [{\citenamefont {Jin}\ \emph {et~al.}(2016)\citenamefont {Jin},
  \citenamefont {Bei},\ and\ \citenamefont {Zhang}}]{Jin2016}%
  \BibitemOpen
  \bibfield  {author} {\bibinfo {author} {\bibfnamefont {K.}~\bibnamefont
  {Jin}}, \bibinfo {author} {\bibfnamefont {H.}~\bibnamefont {Bei}}, \ and\
  \bibinfo {author} {\bibfnamefont {Y.}~\bibnamefont {Zhang}},\ }\bibfield
  {title} {\enquote {\bibinfo {title} {Ion irradiation induced defect evolution
  in ni and ni-based fcc equiatomic binary alloys},}\ }\href {\doibase
  10.1016/j.jnucmat.2015.09.009} {\bibfield  {journal} {\bibinfo  {journal} {J.
  Nucl. Mater.}\ }\textbf {\bibinfo {volume} {471}},\ \bibinfo {pages}
  {193--199} (\bibinfo {year} {2016})}\BibitemShut {NoStop}%
\bibitem [{\citenamefont {Velişa}\ \emph {et~al.}(2017)\citenamefont
  {Velişa}, \citenamefont {Ullah}, \citenamefont {Xue}, \citenamefont {Jin},
  \citenamefont {Crespillo}, \citenamefont {Bei}, \citenamefont {Weber},\ and\
  \citenamefont {Zhang}}]{Velisa2017}%
  \BibitemOpen
  \bibfield  {author} {\bibinfo {author} {\bibfnamefont {G.}~\bibnamefont
  {Velişa}}, \bibinfo {author} {\bibfnamefont {M.~W.}\ \bibnamefont {Ullah}},
  \bibinfo {author} {\bibfnamefont {H.}~\bibnamefont {Xue}}, \bibinfo {author}
  {\bibfnamefont {K.}~\bibnamefont {Jin}}, \bibinfo {author} {\bibfnamefont
  {M.~L.}\ \bibnamefont {Crespillo}}, \bibinfo {author} {\bibfnamefont
  {H.}~\bibnamefont {Bei}}, \bibinfo {author} {\bibfnamefont {W.~J.}\
  \bibnamefont {Weber}}, \ and\ \bibinfo {author} {\bibfnamefont
  {Y.}~\bibnamefont {Zhang}},\ }\bibfield  {title} {\enquote {\bibinfo {title}
  {Irradiation-induced damage evolution in concentrated ni-based alloys},}\
  }\href {\doibase 10.1016/j.actamat.2017.06.002} {\bibfield  {journal}
  {\bibinfo  {journal} {Acta Mater.}\ }\textbf {\bibinfo {volume} {135}},\
  \bibinfo {pages} {54--60} (\bibinfo {year} {2017})}\BibitemShut {NoStop}%
\bibitem [{\citenamefont {Fan}\ \emph {et~al.}(2019)\citenamefont {Fan},
  \citenamefont {Velisa}, \citenamefont {Jin}, \citenamefont {Crespillo},
  \citenamefont {Bei},\ and\ \citenamefont {Weber}}]{Fan2019}%
  \BibitemOpen
  \bibfield  {author} {\bibinfo {author} {\bibfnamefont {Z.}~\bibnamefont
  {Fan}}, \bibinfo {author} {\bibfnamefont {G.}~\bibnamefont {Velisa}},
  \bibinfo {author} {\bibfnamefont {K.}~\bibnamefont {Jin}}, \bibinfo {author}
  {\bibfnamefont {M.~L.}\ \bibnamefont {Crespillo}}, \bibinfo {author}
  {\bibfnamefont {H.}~\bibnamefont {Bei}}, \ and\ \bibinfo {author}
  {\bibfnamefont {W.~J.}\ \bibnamefont {Weber}},\ }\bibfield  {title} {\enquote
  {\bibinfo {title} {Temperature-dependent defect accumulation and evolution in
  {Ni}-irradiated {NiFe} concentrated solid-solution alloy},}\ }\href@noop {}
  {\bibfield  {journal} {\bibinfo  {journal} {J. Nucl. Mater.}\ }\textbf
  {\bibinfo {volume} {519}},\ \bibinfo {pages} {1--9} (\bibinfo {year}
  {2019})}\BibitemShut {NoStop}%
\bibitem [{\citenamefont {Zhang}\ \emph {et~al.}(2016)\citenamefont {Zhang},
  \citenamefont {Nordlund}, \citenamefont {Djurabekova}, \citenamefont {Zhang},
  \citenamefont {Velisa},\ and\ \citenamefont {Wang}}]{Zhang2016}%
  \BibitemOpen
  \bibfield  {author} {\bibinfo {author} {\bibfnamefont {S.}~\bibnamefont
  {Zhang}}, \bibinfo {author} {\bibfnamefont {K.}~\bibnamefont {Nordlund}},
  \bibinfo {author} {\bibfnamefont {F.}~\bibnamefont {Djurabekova}}, \bibinfo
  {author} {\bibfnamefont {Y.}~\bibnamefont {Zhang}}, \bibinfo {author}
  {\bibfnamefont {G.}~\bibnamefont {Velisa}}, \ and\ \bibinfo {author}
  {\bibfnamefont {T.~S.}\ \bibnamefont {Wang}},\ }\bibfield  {title} {\enquote
  {\bibinfo {title} {Simulation of rutherford backscattering spectrometry from
  arbitrary atom structures},}\ }\href {\doibase 10.1103/physreve.94.043319}
  {\bibfield  {journal} {\bibinfo  {journal} {Phys. Rev. E}\ }\textbf {\bibinfo
  {volume} {94}},\ \bibinfo {pages} {043319} (\bibinfo {year}
  {2016})}\BibitemShut {NoStop}%
\bibitem [{\citenamefont {Zhang}\ \emph {et~al.}(2017)\citenamefont {Zhang},
  \citenamefont {Nordlund}, \citenamefont {Djurabekova}, \citenamefont
  {Granberg}, \citenamefont {Zhang},\ and\ \citenamefont {Wang}}]{Zhang2017}%
  \BibitemOpen
  \bibfield  {author} {\bibinfo {author} {\bibfnamefont {S.}~\bibnamefont
  {Zhang}}, \bibinfo {author} {\bibfnamefont {K.}~\bibnamefont {Nordlund}},
  \bibinfo {author} {\bibfnamefont {F.}~\bibnamefont {Djurabekova}}, \bibinfo
  {author} {\bibfnamefont {F.}~\bibnamefont {Granberg}}, \bibinfo {author}
  {\bibfnamefont {Y.}~\bibnamefont {Zhang}}, \ and\ \bibinfo {author}
  {\bibfnamefont {T.~S.}\ \bibnamefont {Wang}},\ }\bibfield  {title} {\enquote
  {\bibinfo {title} {Radiation damage buildup by athermal defect reactions in
  nickel and concentrated nickel alloys},}\ }\href {\doibase
  10.1080/21663831.2017.1311284} {\bibfield  {journal} {\bibinfo  {journal}
  {Mater. Res. Lett.}\ }\textbf {\bibinfo {volume} {5}},\ \bibinfo {pages}
  {433--439} (\bibinfo {year} {2017})}\BibitemShut {NoStop}%
\bibitem [{\citenamefont {Levo}\ \emph {et~al.}(2021)\citenamefont {Levo},
  \citenamefont {Granberg}, \citenamefont {Nordlund},\ and\ \citenamefont
  {Djurabekova}}]{Levo2021}%
  \BibitemOpen
  \bibfield  {author} {\bibinfo {author} {\bibfnamefont {E.}~\bibnamefont
  {Levo}}, \bibinfo {author} {\bibfnamefont {F.}~\bibnamefont {Granberg}},
  \bibinfo {author} {\bibfnamefont {K.}~\bibnamefont {Nordlund}}, \ and\
  \bibinfo {author} {\bibfnamefont {F.}~\bibnamefont {Djurabekova}},\
  }\bibfield  {title} {\enquote {\bibinfo {title} {Temperature effect on
  irradiation damage in equiatomic multi-component alloys},}\ }\href@noop {}
  {\bibfield  {journal} {\bibinfo  {journal} {Comput. Mater. Sci.}\ }\textbf
  {\bibinfo {volume} {197}},\ \bibinfo {pages} {110571} (\bibinfo {year}
  {2021})}\BibitemShut {NoStop}%
\bibitem [{\citenamefont {Short}\ \emph {et~al.}(2015)\citenamefont {Short},
  \citenamefont {Dennett}, \citenamefont {Ferry}, \citenamefont {Yang},
  \citenamefont {Mishra}, \citenamefont {Eliason}, \citenamefont {Vega-Flick},
  \citenamefont {Maznev},\ and\ \citenamefont {Nelson}}]{Short2015}%
  \BibitemOpen
  \bibfield  {author} {\bibinfo {author} {\bibfnamefont {M.~P.}\ \bibnamefont
  {Short}}, \bibinfo {author} {\bibfnamefont {C.~A.}\ \bibnamefont {Dennett}},
  \bibinfo {author} {\bibfnamefont {S.~E.}\ \bibnamefont {Ferry}}, \bibinfo
  {author} {\bibfnamefont {Y.}~\bibnamefont {Yang}}, \bibinfo {author}
  {\bibfnamefont {V.~K.}\ \bibnamefont {Mishra}}, \bibinfo {author}
  {\bibfnamefont {J.~K.}\ \bibnamefont {Eliason}}, \bibinfo {author}
  {\bibfnamefont {A.}~\bibnamefont {Vega-Flick}}, \bibinfo {author}
  {\bibfnamefont {A.~A.}\ \bibnamefont {Maznev}}, \ and\ \bibinfo {author}
  {\bibfnamefont {K.~A.}\ \bibnamefont {Nelson}},\ }\bibfield  {title}
  {\enquote {\bibinfo {title} {Applications of transient grating spectroscopy
  to radiation materials science},}\ }\href@noop {} {\bibfield  {journal}
  {\bibinfo  {journal} {JOM}\ }\textbf {\bibinfo {volume} {67}},\ \bibinfo
  {pages} {1840--1848} (\bibinfo {year} {2015})}\BibitemShut {NoStop}%
\bibitem [{\citenamefont {Dennett}\ \emph {et~al.}(2016)\citenamefont
  {Dennett}, \citenamefont {Cao}, \citenamefont {Ferry}, \citenamefont
  {Vega-Flick}, \citenamefont {Maznev}, \citenamefont {Nelson}, \citenamefont
  {Every},\ and\ \citenamefont {Short}}]{Dennett2016}%
  \BibitemOpen
  \bibfield  {author} {\bibinfo {author} {\bibfnamefont {C.~A.}\ \bibnamefont
  {Dennett}}, \bibinfo {author} {\bibfnamefont {P.}~\bibnamefont {Cao}},
  \bibinfo {author} {\bibfnamefont {S.~E.}\ \bibnamefont {Ferry}}, \bibinfo
  {author} {\bibfnamefont {A.}~\bibnamefont {Vega-Flick}}, \bibinfo {author}
  {\bibfnamefont {A.~A.}\ \bibnamefont {Maznev}}, \bibinfo {author}
  {\bibfnamefont {K.~A.}\ \bibnamefont {Nelson}}, \bibinfo {author}
  {\bibfnamefont {A.~G.}\ \bibnamefont {Every}}, \ and\ \bibinfo {author}
  {\bibfnamefont {M.~P.}\ \bibnamefont {Short}},\ }\bibfield  {title} {\enquote
  {\bibinfo {title} {Bridging the gap to mesoscale radiation materials science
  with transient grating spectroscopy},}\ }\href {\doibase
  10.1103/PhysRevB.94.214106} {\bibfield  {journal} {\bibinfo  {journal} {Phys.
  Rev. B}\ }\textbf {\bibinfo {volume} {94}},\ \bibinfo {pages} {214106}
  (\bibinfo {year} {2016})}\BibitemShut {NoStop}%
\bibitem [{\citenamefont {Hofmann}\ \emph {et~al.}(2019)\citenamefont
  {Hofmann}, \citenamefont {Short},\ and\ \citenamefont
  {Dennett}}]{Hofmann2019}%
  \BibitemOpen
  \bibfield  {author} {\bibinfo {author} {\bibfnamefont {F.}~\bibnamefont
  {Hofmann}}, \bibinfo {author} {\bibfnamefont {M.~P.}\ \bibnamefont {Short}},
  \ and\ \bibinfo {author} {\bibfnamefont {C.~A.}\ \bibnamefont {Dennett}},\
  }\bibfield  {title} {\enquote {\bibinfo {title} {Transient grating
  spectroscopy: An ultrarapid, nondestructive materials evaluation
  technique},}\ }\href {\doibase 10.1557/mrs.2019.104} {\bibfield  {journal}
  {\bibinfo  {journal} {MRS Bulletin}\ }\textbf {\bibinfo {volume} {44}},\
  \bibinfo {pages} {392--402} (\bibinfo {year} {2019})}\BibitemShut {NoStop}%
\bibitem [{\citenamefont {K\"{a}ding}\ \emph {et~al.}(1995)\citenamefont
  {K\"{a}ding}, \citenamefont {Skurk}, \citenamefont {Maznev},\ and\
  \citenamefont {Matthias}}]{Kading1995}%
  \BibitemOpen
  \bibfield  {author} {\bibinfo {author} {\bibfnamefont {O.~W.}\ \bibnamefont
  {K\"{a}ding}}, \bibinfo {author} {\bibfnamefont {H.}~\bibnamefont {Skurk}},
  \bibinfo {author} {\bibfnamefont {A.~A.}\ \bibnamefont {Maznev}}, \ and\
  \bibinfo {author} {\bibfnamefont {E.}~\bibnamefont {Matthias}},\ }\bibfield
  {title} {\enquote {\bibinfo {title} {Transient thermal gratings at surfaces
  for thermal characterization of bulk materials and thin films},}\ }\href
  {\doibase 10.1007/BF01538190} {\bibfield  {journal} {\bibinfo  {journal}
  {Appl. Phys. A}\ }\textbf {\bibinfo {volume} {61}},\ \bibinfo {pages}
  {253--261} (\bibinfo {year} {1995})}\BibitemShut {NoStop}%
\bibitem [{\citenamefont {Johnson}\ \emph {et~al.}(2012)\citenamefont
  {Johnson}, \citenamefont {Maznev}, \citenamefont {Bulsara}, \citenamefont
  {Fitzgerald}, \citenamefont {Harman}, \citenamefont {Calawa}, \citenamefont
  {Vineis}, \citenamefont {Turner},\ and\ \citenamefont
  {Nelson}}]{Johnson2012}%
  \BibitemOpen
  \bibfield  {author} {\bibinfo {author} {\bibfnamefont {J.~A.}\ \bibnamefont
  {Johnson}}, \bibinfo {author} {\bibfnamefont {A.~A.}\ \bibnamefont {Maznev}},
  \bibinfo {author} {\bibfnamefont {M.~T.}\ \bibnamefont {Bulsara}}, \bibinfo
  {author} {\bibfnamefont {E.~A.}\ \bibnamefont {Fitzgerald}}, \bibinfo
  {author} {\bibfnamefont {T.~C.}\ \bibnamefont {Harman}}, \bibinfo {author}
  {\bibfnamefont {S.}~\bibnamefont {Calawa}}, \bibinfo {author} {\bibfnamefont
  {C.~J.}\ \bibnamefont {Vineis}}, \bibinfo {author} {\bibfnamefont
  {G.}~\bibnamefont {Turner}}, \ and\ \bibinfo {author} {\bibfnamefont {K.~A.}\
  \bibnamefont {Nelson}},\ }\bibfield  {title} {\enquote {\bibinfo {title}
  {Phase-controlled, heterodyne laser-induced transient grating measurements of
  thermal transport properties in opaque material},}\ }\href {\doibase
  10.1063/1.3675467} {\bibfield  {journal} {\bibinfo  {journal} {J. Appl.
  Phys.}\ }\textbf {\bibinfo {volume} {111}},\ \bibinfo {pages} {023503}
  (\bibinfo {year} {2012})}\BibitemShut {NoStop}%
\bibitem [{\citenamefont {Dennett}\ and\ \citenamefont
  {Short}(2018)}]{Dennett2018a}%
  \BibitemOpen
  \bibfield  {author} {\bibinfo {author} {\bibfnamefont {C.~A.}\ \bibnamefont
  {Dennett}}\ and\ \bibinfo {author} {\bibfnamefont {M.~P.}\ \bibnamefont
  {Short}},\ }\bibfield  {title} {\enquote {\bibinfo {title} {Thermal
  diffusivity determination using heterodyne phase insensitive transient
  grating spectroscopy},}\ }\href {\doibase https://doi.org/10.1063/1.5026429}
  {\bibfield  {journal} {\bibinfo  {journal} {J. Appl. Phys.}\ }\textbf
  {\bibinfo {volume} {123}},\ \bibinfo {pages} {215109} (\bibinfo {year}
  {2018})}\BibitemShut {NoStop}%
\bibitem [{\citenamefont {Dennett}\ \emph {et~al.}(2018)\citenamefont
  {Dennett}, \citenamefont {So}, \citenamefont {Kushima}, \citenamefont
  {Buller}, \citenamefont {Hattar},\ and\ \citenamefont {Short}}]{Dennett2018}%
  \BibitemOpen
  \bibfield  {author} {\bibinfo {author} {\bibfnamefont {C.~A.}\ \bibnamefont
  {Dennett}}, \bibinfo {author} {\bibfnamefont {K.~P.}\ \bibnamefont {So}},
  \bibinfo {author} {\bibfnamefont {A.}~\bibnamefont {Kushima}}, \bibinfo
  {author} {\bibfnamefont {D.~L.}\ \bibnamefont {Buller}}, \bibinfo {author}
  {\bibfnamefont {K.}~\bibnamefont {Hattar}}, \ and\ \bibinfo {author}
  {\bibfnamefont {M.~P.}\ \bibnamefont {Short}},\ }\bibfield  {title} {\enquote
  {\bibinfo {title} {Detecting self-ion irradiation-induced void swelling in
  pure copper using transient grating spectroscopy},}\ }\href {\doibase
  10.1016/j.actamat.2017.12.007} {\bibfield  {journal} {\bibinfo  {journal}
  {Acta Mater.}\ }\textbf {\bibinfo {volume} {145}},\ \bibinfo {pages}
  {496--503} (\bibinfo {year} {2018})}\BibitemShut {NoStop}%
\bibitem [{\citenamefont {AlMousa}\ \emph {et~al.}(2021)\citenamefont
  {AlMousa}, \citenamefont {Dacus}, \citenamefont {Woller}, \citenamefont
  {Shin}, \citenamefont {Jang}, \citenamefont {Shao}, \citenamefont {Garner},
  \citenamefont {Gabriel},\ and\ \citenamefont {Short}}]{AlMousa2021}%
  \BibitemOpen
  \bibfield  {author} {\bibinfo {author} {\bibfnamefont {N.}~\bibnamefont
  {AlMousa}}, \bibinfo {author} {\bibfnamefont {B.~R.}\ \bibnamefont {Dacus}},
  \bibinfo {author} {\bibfnamefont {K.~B.}\ \bibnamefont {Woller}}, \bibinfo
  {author} {\bibfnamefont {J.~H.}\ \bibnamefont {Shin}}, \bibinfo {author}
  {\bibfnamefont {C.}~\bibnamefont {Jang}}, \bibinfo {author} {\bibfnamefont
  {L.}~\bibnamefont {Shao}}, \bibinfo {author} {\bibfnamefont {F.~A.}\
  \bibnamefont {Garner}}, \bibinfo {author} {\bibfnamefont {A.}~\bibnamefont
  {Gabriel}}, \ and\ \bibinfo {author} {\bibfnamefont {M.~P.}\ \bibnamefont
  {Short}},\ }\bibfield  {title} {\enquote {\bibinfo {title} {On the use of
  non-destructive, gigahertz ultrasonics to rapidly screen irradiated steels
  for swelling resistance},}\ }\href {\doibase
  https://doi.org/10.1016/j.matchar.2021.111017} {\bibfield  {journal}
  {\bibinfo  {journal} {Mater. Charact.}\ }\textbf {\bibinfo {volume} {174}},\
  \bibinfo {pages} {111017} (\bibinfo {year} {2021})}\BibitemShut {NoStop}%
\bibitem [{\citenamefont {Hofmann}\ \emph
  {et~al.}(2015{\natexlab{a}})\citenamefont {Hofmann}, \citenamefont
  {Nguyen-Manh}, \citenamefont {Gilbert}, \citenamefont {Beck}, \citenamefont
  {Eliason}, \citenamefont {Maznev}, \citenamefont {Liu}, \citenamefont
  {Armstrong}, \citenamefont {Nelson},\ and\ \citenamefont
  {Dudarev}}]{Hofmann2015}%
  \BibitemOpen
  \bibfield  {author} {\bibinfo {author} {\bibfnamefont {F.}~\bibnamefont
  {Hofmann}}, \bibinfo {author} {\bibfnamefont {D.}~\bibnamefont
  {Nguyen-Manh}}, \bibinfo {author} {\bibfnamefont {M.~R.}\ \bibnamefont
  {Gilbert}}, \bibinfo {author} {\bibfnamefont {C.~E.}\ \bibnamefont {Beck}},
  \bibinfo {author} {\bibfnamefont {J.~K.}\ \bibnamefont {Eliason}}, \bibinfo
  {author} {\bibfnamefont {A.~A.}\ \bibnamefont {Maznev}}, \bibinfo {author}
  {\bibfnamefont {W.}~\bibnamefont {Liu}}, \bibinfo {author} {\bibfnamefont
  {D.~E.~J.}\ \bibnamefont {Armstrong}}, \bibinfo {author} {\bibfnamefont
  {K.~A.}\ \bibnamefont {Nelson}}, \ and\ \bibinfo {author} {\bibfnamefont
  {S.~L.}\ \bibnamefont {Dudarev}},\ }\bibfield  {title} {\enquote {\bibinfo
  {title} {Lattice swelling and modulus change in a helium-implanted tungsten
  alloy: {X}-ray micro-diffraction, surface acoustic wave measurements, and
  multiscale modelling},}\ }\href {\doibase
  http://dx.doi.org/10.1016/j.actamat.2015.01.055} {\bibfield  {journal}
  {\bibinfo  {journal} {Acta Mater.}\ }\textbf {\bibinfo {volume} {89}},\
  \bibinfo {pages} {352--363} (\bibinfo {year}
  {2015}{\natexlab{a}})}\BibitemShut {NoStop}%
\bibitem [{\citenamefont {Hofmann}\ \emph
  {et~al.}(2015{\natexlab{b}})\citenamefont {Hofmann}, \citenamefont {Mason},
  \citenamefont {Eliason}, \citenamefont {Maznev}, \citenamefont {Nelson},\
  and\ \citenamefont {Dudarev}}]{Hofmann2015a}%
  \BibitemOpen
  \bibfield  {author} {\bibinfo {author} {\bibfnamefont {F.}~\bibnamefont
  {Hofmann}}, \bibinfo {author} {\bibfnamefont {D.~R.}\ \bibnamefont {Mason}},
  \bibinfo {author} {\bibfnamefont {J.~K.}\ \bibnamefont {Eliason}}, \bibinfo
  {author} {\bibfnamefont {A.~A.}\ \bibnamefont {Maznev}}, \bibinfo {author}
  {\bibfnamefont {K.~A.}\ \bibnamefont {Nelson}}, \ and\ \bibinfo {author}
  {\bibfnamefont {S.~L.}\ \bibnamefont {Dudarev}},\ }\bibfield  {title}
  {\enquote {\bibinfo {title} {Non-contact measurement of thermal diffusivity
  in ion-implanted nuclear materials},}\ }\href@noop {} {\bibfield  {journal}
  {\bibinfo  {journal} {Sci. Rep.}\ }\textbf {\bibinfo {volume} {5}},\ \bibinfo
  {pages} {16042} (\bibinfo {year} {2015}{\natexlab{b}})}\BibitemShut {NoStop}%
\bibitem [{\citenamefont {Dennett}\ \emph {et~al.}(2019)\citenamefont
  {Dennett}, \citenamefont {Buller}, \citenamefont {Hattar},\ and\
  \citenamefont {Short}}]{Dennett2019}%
  \BibitemOpen
  \bibfield  {author} {\bibinfo {author} {\bibfnamefont {C.~A.}\ \bibnamefont
  {Dennett}}, \bibinfo {author} {\bibfnamefont {D.~L.}\ \bibnamefont {Buller}},
  \bibinfo {author} {\bibfnamefont {K.}~\bibnamefont {Hattar}}, \ and\ \bibinfo
  {author} {\bibfnamefont {M.~P.}\ \bibnamefont {Short}},\ }\bibfield  {title}
  {\enquote {\bibinfo {title} {Real-time thermomechanical property monitoring
  during ion beam irradiation using \emph{in~situ} transient grating
  spectroscopy},}\ }\href@noop {} {\bibfield  {journal} {\bibinfo  {journal}
  {Nucl. Instrum. Meth. Phys. Res. B}\ }\textbf {\bibinfo {volume} {440}},\
  \bibinfo {pages} {126--138} (\bibinfo {year} {2019})}\BibitemShut {NoStop}%
\bibitem [{\citenamefont {Ferry}\ \emph {et~al.}(2019)\citenamefont {Ferry},
  \citenamefont {Dennett}, \citenamefont {Woller},\ and\ \citenamefont
  {Short}}]{Ferry2019}%
  \BibitemOpen
  \bibfield  {author} {\bibinfo {author} {\bibfnamefont {S.~E.}\ \bibnamefont
  {Ferry}}, \bibinfo {author} {\bibfnamefont {C.~A.}\ \bibnamefont {Dennett}},
  \bibinfo {author} {\bibfnamefont {K.~B.}\ \bibnamefont {Woller}}, \ and\
  \bibinfo {author} {\bibfnamefont {M.~P.}\ \bibnamefont {Short}},\ }\bibfield
  {title} {\enquote {\bibinfo {title} {Inferring radiation-induced
  microstructural evolution in single-crystal niobium through changes in
  thermal transport},}\ }\href {\doibase
  https://doi.org/10.1016/j.jnucmat.2019.06.015} {\bibfield  {journal}
  {\bibinfo  {journal} {J. Nucl. Mater.}\ }\textbf {\bibinfo {volume} {523}},\
  \bibinfo {pages} {378--382} (\bibinfo {year} {2019})}\BibitemShut {NoStop}%
\bibitem [{\citenamefont {Dennett}\ \emph {et~al.}(2021)\citenamefont
  {Dennett}, \citenamefont {Dacus}, \citenamefont {Barr}, \citenamefont
  {Clark}, \citenamefont {Bei}, \citenamefont {Zhang}, \citenamefont {Short},\
  and\ \citenamefont {Hattar}}]{Dennett2021}%
  \BibitemOpen
  \bibfield  {author} {\bibinfo {author} {\bibfnamefont {C.~A.}\ \bibnamefont
  {Dennett}}, \bibinfo {author} {\bibfnamefont {B.~R.}\ \bibnamefont {Dacus}},
  \bibinfo {author} {\bibfnamefont {C.~M.}\ \bibnamefont {Barr}}, \bibinfo
  {author} {\bibfnamefont {T.}~\bibnamefont {Clark}}, \bibinfo {author}
  {\bibfnamefont {H.}~\bibnamefont {Bei}}, \bibinfo {author} {\bibfnamefont
  {Y.}~\bibnamefont {Zhang}}, \bibinfo {author} {\bibfnamefont {M.~P.}\
  \bibnamefont {Short}}, \ and\ \bibinfo {author} {\bibfnamefont
  {K.}~\bibnamefont {Hattar}},\ }\bibfield  {title} {\enquote {\bibinfo {title}
  {The dynamic evolution of swelling in nickel concentrated solid solution
  alloys through \emph{in~situ} property monitoring},}\ }\href {\doibase
  https://doi.org/10.1016/j.apmt.2021.101187} {\bibfield  {journal} {\bibinfo
  {journal} {Appl. Mater. Today}\ }\textbf {\bibinfo {volume} {25}},\ \bibinfo
  {pages} {101187} (\bibinfo {year} {2021})}\BibitemShut {NoStop}%
\bibitem [{\citenamefont {Schilling}(1978)}]{Schilling1978}%
  \BibitemOpen
  \bibfield  {author} {\bibinfo {author} {\bibfnamefont {W.}~\bibnamefont
  {Schilling}},\ }\bibfield  {title} {\enquote {\bibinfo {title} {Radiation
  induced damage in metals},}\ }\href {\doibase 10.1016/0022-3115(78)90384-7}
  {\bibfield  {journal} {\bibinfo  {journal} {J. Nucl. Mater.}\ }\textbf
  {\bibinfo {volume} {72}},\ \bibinfo {pages} {1--4} (\bibinfo {year}
  {1978})}\BibitemShut {NoStop}%
\bibitem [{\citenamefont {Snead}\ \emph {et~al.}(2019)\citenamefont {Snead},
  \citenamefont {Katoh}, \citenamefont {Koyanagi},\ and\ \citenamefont
  {Terrani}}]{Snead2019}%
  \BibitemOpen
  \bibfield  {author} {\bibinfo {author} {\bibfnamefont {L.~L.}\ \bibnamefont
  {Snead}}, \bibinfo {author} {\bibfnamefont {Y.}~\bibnamefont {Katoh}},
  \bibinfo {author} {\bibfnamefont {T.}~\bibnamefont {Koyanagi}}, \ and\
  \bibinfo {author} {\bibfnamefont {K.}~\bibnamefont {Terrani}},\ }\bibfield
  {title} {\enquote {\bibinfo {title} {Stored energy release in neutron
  irradiated silicon carbide},}\ }\href {\doibase
  10.1016/j.jnucmat.2018.12.005} {\bibfield  {journal} {\bibinfo  {journal} {J.
  Nucl. Mater.}\ }\textbf {\bibinfo {volume} {514}},\ \bibinfo {pages}
  {181--188} (\bibinfo {year} {2019})}\BibitemShut {NoStop}%
\bibitem [{\citenamefont {Iwata}(1985)}]{Iwata1985}%
  \BibitemOpen
  \bibfield  {author} {\bibinfo {author} {\bibfnamefont {T.}~\bibnamefont
  {Iwata}},\ }\bibfield  {title} {\enquote {\bibinfo {title} {Fine structure of
  {W}igner energy release spectrum in neutron irradiated graphite},}\ }\href
  {\doibase 10.1016/0022-3115(85)90168-0} {\bibfield  {journal} {\bibinfo
  {journal} {J. Nucl. Mater.}\ }\textbf {\bibinfo {volume} {133--134}},\
  \bibinfo {pages} {361--364} (\bibinfo {year} {1985})}\BibitemShut {NoStop}%
\bibitem [{\citenamefont {Staicu}\ \emph {et~al.}(2010)\citenamefont {Staicu},
  \citenamefont {Wiss}, \citenamefont {Rondinella}, \citenamefont {Hiernaut},
  \citenamefont {Konings},\ and\ \citenamefont {Ronchi}}]{Staicu2010}%
  \BibitemOpen
  \bibfield  {author} {\bibinfo {author} {\bibfnamefont {D.}~\bibnamefont
  {Staicu}}, \bibinfo {author} {\bibfnamefont {T.}~\bibnamefont {Wiss}},
  \bibinfo {author} {\bibfnamefont {V.~V.}\ \bibnamefont {Rondinella}},
  \bibinfo {author} {\bibfnamefont {J.~P.}\ \bibnamefont {Hiernaut}}, \bibinfo
  {author} {\bibfnamefont {R.~J.~M.}\ \bibnamefont {Konings}}, \ and\ \bibinfo
  {author} {\bibfnamefont {C.}~\bibnamefont {Ronchi}},\ }\bibfield  {title}
  {\enquote {\bibinfo {title} {Impact of auto-irradiation on the thermophysical
  properties of oxide nuclear reactor fuels},}\ }\href {\doibase
  10.1016/j.jnucmat.2009.11.024} {\bibfield  {journal} {\bibinfo  {journal} {J.
  Nucl. Mater.}\ }\textbf {\bibinfo {volume} {397}},\ \bibinfo {pages} {8--18}
  (\bibinfo {year} {2010})}\BibitemShut {NoStop}%
\bibitem [{\citenamefont {Vainshtein}\ and\ \citenamefont
  {Hartog}(2000)}]{Vainshtein2000}%
  \BibitemOpen
  \bibfield  {author} {\bibinfo {author} {\bibfnamefont {D.~I.}\ \bibnamefont
  {Vainshtein}}\ and\ \bibinfo {author} {\bibfnamefont {H.~W.~Den}\
  \bibnamefont {Hartog}},\ }\bibfield  {title} {\enquote {\bibinfo {title}
  {Explosive decomposition of heavily irradiated {NaCl}},}\ }\href {\doibase
  10.1080/10420150008211811} {\bibfield  {journal} {\bibinfo  {journal}
  {Radiat. Eff. Defects Solids}\ }\textbf {\bibinfo {volume} {152}},\ \bibinfo
  {pages} {23--37} (\bibinfo {year} {2000})}\BibitemShut {NoStop}%
\bibitem [{\citenamefont {Wigner}(1942)}]{Wigner1942}%
  \BibitemOpen
  \bibfield  {author} {\bibinfo {author} {\bibfnamefont {E.}~\bibnamefont
  {Wigner}},\ }\href@noop {} {\emph {\bibinfo {title} {Report for Month Ending
  December 15}}},\ \bibinfo {type} {Tech. Rep.}\ (\bibinfo  {institution} {US
  Atomic Energy Commission Report CP-387, Univ. of Chicago, Chicago},\ \bibinfo
  {year} {1942})\BibitemShut {NoStop}%
\bibitem [{\citenamefont {Blewitt}\ \emph {et~al.}(1959)\citenamefont
  {Blewitt}, \citenamefont {Coltman},\ and\ \citenamefont
  {Klabunde}}]{Blewitt1959}%
  \BibitemOpen
  \bibfield  {author} {\bibinfo {author} {\bibfnamefont {T.~H.}\ \bibnamefont
  {Blewitt}}, \bibinfo {author} {\bibfnamefont {R.~R.}\ \bibnamefont
  {Coltman}}, \ and\ \bibinfo {author} {\bibfnamefont {C.~E.}\ \bibnamefont
  {Klabunde}},\ }\bibfield  {title} {\enquote {\bibinfo {title} {Stored energy
  of irradiated copper},}\ }\href {\doibase 10.1103/physrevlett.3.132}
  {\bibfield  {journal} {\bibinfo  {journal} {Phys. Rev. Lett.}\ }\textbf
  {\bibinfo {volume} {3}},\ \bibinfo {pages} {132--134} (\bibinfo {year}
  {1959})}\BibitemShut {NoStop}%
\bibitem [{\citenamefont {Richard}\ \emph {et~al.}(1990)\citenamefont
  {Richard}, \citenamefont {Chaplin}, \citenamefont {Coltman}, \citenamefont
  {Kerchner},\ and\ \citenamefont {Klabunde}}]{Richard1990}%
  \BibitemOpen
  \bibfield  {author} {\bibinfo {author} {\bibfnamefont {R.~T.}\ \bibnamefont
  {Richard}}, \bibinfo {author} {\bibfnamefont {R.~L.}\ \bibnamefont
  {Chaplin}}, \bibinfo {author} {\bibfnamefont {R.~R.}\ \bibnamefont
  {Coltman}}, \bibinfo {author} {\bibfnamefont {H.~R.}\ \bibnamefont
  {Kerchner}}, \ and\ \bibinfo {author} {\bibfnamefont {C.~E.}\ \bibnamefont
  {Klabunde}},\ }\bibfield  {title} {\enquote {\bibinfo {title} {Stored energy
  recovery of irradiated copper},}\ }\href {\doibase 10.1080/10420159008213042}
  {\bibfield  {journal} {\bibinfo  {journal} {Radiat. Eff. Defects Solids}\
  }\textbf {\bibinfo {volume} {112}},\ \bibinfo {pages} {161--179} (\bibinfo
  {year} {1990})}\BibitemShut {NoStop}%
\bibitem [{\citenamefont {Pedchenko}\ and\ \citenamefont
  {Karasev}(1971)}]{Pedchenko1971}%
  \BibitemOpen
  \bibfield  {author} {\bibinfo {author} {\bibfnamefont {K.~S.}\ \bibnamefont
  {Pedchenko}}\ and\ \bibinfo {author} {\bibfnamefont {V.~S.}\ \bibnamefont
  {Karasev}},\ }\bibfield  {title} {\enquote {\bibinfo {title} {Stored energy
  in neutron-bombarded metals},}\ }\href {https://doi.org/10.1007/BF00722627}
  {\bibfield  {journal} {\bibinfo  {journal} {Transl. of Fiziko-khimicheskaya
  mekhanika materialov / Academy of Sciences of the Ukrainian SSR}\ }\textbf
  {\bibinfo {volume} {4}},\ \bibinfo {pages} {332--336} (\bibinfo {year}
  {1971})}\BibitemShut {NoStop}%
\bibitem [{\citenamefont {Lambri}\ \emph {et~al.}(2009)\citenamefont {Lambri},
  \citenamefont {Zelada-Lambri}, \citenamefont {Cuello}, \citenamefont
  {Bozzano},\ and\ \citenamefont {García}}]{Lambri2009}%
  \BibitemOpen
  \bibfield  {author} {\bibinfo {author} {\bibfnamefont {O.~A.}\ \bibnamefont
  {Lambri}}, \bibinfo {author} {\bibfnamefont {G.~I.}\ \bibnamefont
  {Zelada-Lambri}}, \bibinfo {author} {\bibfnamefont {G.~J.}\ \bibnamefont
  {Cuello}}, \bibinfo {author} {\bibfnamefont {P.~B.}\ \bibnamefont {Bozzano}},
  \ and\ \bibinfo {author} {\bibfnamefont {J.~A.}\ \bibnamefont {García}},\
  }\bibfield  {title} {\enquote {\bibinfo {title} {Study of the temperature
  evolution of defect agglomerates in neutron irradiated molybdenum single
  crystals},}\ }\href {\doibase 10.1016/j.jnucmat.2008.12.312} {\bibfield
  {journal} {\bibinfo  {journal} {J. Nucl. Mater.}\ }\textbf {\bibinfo {volume}
  {385}},\ \bibinfo {pages} {552--558} (\bibinfo {year} {2009})}\BibitemShut
  {NoStop}%
\bibitem [{\citenamefont {Blewitt}\ \emph {et~al.}(1961)\citenamefont
  {Blewitt}, \citenamefont {Sekula},\ and\ \citenamefont
  {Diehl}}]{Blewitt1961}%
  \BibitemOpen
  \bibfield  {author} {\bibinfo {author} {\bibfnamefont {T.~H.}\ \bibnamefont
  {Blewitt}}, \bibinfo {author} {\bibfnamefont {S.~T.}\ \bibnamefont {Sekula}},
  \ and\ \bibinfo {author} {\bibfnamefont {J.}~\bibnamefont {Diehl}},\
  }\bibfield  {title} {\enquote {\bibinfo {title} {Energy release in
  reactor-irradiated copper. {II}. 600$^\circ$ to 700$^\circ${K} release},}\
  }\href {\doibase 10.1103/physrev.122.53} {\bibfield  {journal} {\bibinfo
  {journal} {Phys. Rev.}\ }\textbf {\bibinfo {volume} {122}},\ \bibinfo {pages}
  {53--57} (\bibinfo {year} {1961})}\BibitemShut {NoStop}%
\bibitem [{\citenamefont {Toktogulova}\ \emph {et~al.}(2010)\citenamefont
  {Toktogulova}, \citenamefont {Gusev}, \citenamefont {Maksimkin},\ and\
  \citenamefont {Garner}}]{Toktogulova2010}%
  \BibitemOpen
  \bibfield  {author} {\bibinfo {author} {\bibfnamefont {D.~A.}\ \bibnamefont
  {Toktogulova}}, \bibinfo {author} {\bibfnamefont {M.~N.}\ \bibnamefont
  {Gusev}}, \bibinfo {author} {\bibfnamefont {O.~P.}\ \bibnamefont
  {Maksimkin}}, \ and\ \bibinfo {author} {\bibfnamefont {F.~A.}\ \bibnamefont
  {Garner}},\ }\bibfield  {title} {\enquote {\bibinfo {title} {Influence of
  neutron irradiation on energy accumulation and dissipation during plastic
  flow and hardening of metallic polycrystals},}\ }\href {\doibase
  10.1520/stp49013s} {\bibfield  {journal} {\bibinfo  {journal} {ASTM STP 1513
  On Effects of Radiation}\ ,\ \bibinfo {pages} {238--238--14}} (\bibinfo
  {year} {2010})}\BibitemShut {NoStop}%
\bibitem [{\citenamefont {Ennaceur}\ and\ \citenamefont
  {Migliori}(2018)}]{Ennaceur2018}%
  \BibitemOpen
  \bibfield  {author} {\bibinfo {author} {\bibfnamefont {S.~M.}\ \bibnamefont
  {Ennaceur}}\ and\ \bibinfo {author} {\bibfnamefont {A.}~\bibnamefont
  {Migliori}},\ }\bibfield  {title} {\enquote {\bibinfo {title} {Toward an
  understanding of aging in plutonium from direct measurements of stored
  energy},}\ }\href {\doibase 10.1080/09500839.2019.1583392} {\bibfield
  {journal} {\bibinfo  {journal} {Phil. Mag.}\ }\textbf {\bibinfo {volume}
  {98}},\ \bibinfo {pages} {502--510} (\bibinfo {year} {2018})}\BibitemShut
  {NoStop}%
\bibitem [{\citenamefont {Lee}\ \emph {et~al.}(2007)\citenamefont {Lee},
  \citenamefont {Cheon}, \citenamefont {Koo}, \citenamefont {Oh}, \citenamefont
  {Yim}, \citenamefont {Sohn}, \citenamefont {Baryshnikov},\ and\ \citenamefont
  {Gaiduchenko}}]{Lee2007}%
  \BibitemOpen
  \bibfield  {author} {\bibinfo {author} {\bibfnamefont {B.}~\bibnamefont
  {Lee}}, \bibinfo {author} {\bibfnamefont {J.-S.}\ \bibnamefont {Cheon}},
  \bibinfo {author} {\bibfnamefont {Y.-H.}\ \bibnamefont {Koo}}, \bibinfo
  {author} {\bibfnamefont {J.-Y.}\ \bibnamefont {Oh}}, \bibinfo {author}
  {\bibfnamefont {J.-S.}\ \bibnamefont {Yim}}, \bibinfo {author} {\bibfnamefont
  {D.-S.}\ \bibnamefont {Sohn}}, \bibinfo {author} {\bibfnamefont
  {M.}~\bibnamefont {Baryshnikov}}, \ and\ \bibinfo {author} {\bibfnamefont
  {A.}~\bibnamefont {Gaiduchenko}},\ }\bibfield  {title} {\enquote {\bibinfo
  {title} {Measurement of the specific heat of {Zr}–40wt\%{U} metallic
  fuel},}\ }\href {\doibase 10.1016/j.jnucmat.2006.10.023} {\bibfield
  {journal} {\bibinfo  {journal} {J. Nucl. Mater.}\ }\textbf {\bibinfo {volume}
  {360}},\ \bibinfo {pages} {315--320} (\bibinfo {year} {2007})}\BibitemShut
  {NoStop}%
\bibitem [{\citenamefont {Béland}\ \emph {et~al.}(2013)\citenamefont
  {Béland}, \citenamefont {Anahory}, \citenamefont {Smeets}, \citenamefont
  {Guihard}, \citenamefont {Brommer}, \citenamefont {Joly}, \citenamefont
  {Pothier}, \citenamefont {Lewis}, \citenamefont {Mousseau},\ and\
  \citenamefont {Schiettekatte}}]{Beland2013}%
  \BibitemOpen
  \bibfield  {author} {\bibinfo {author} {\bibfnamefont {L.~K.}\ \bibnamefont
  {Béland}}, \bibinfo {author} {\bibfnamefont {Y.}~\bibnamefont {Anahory}},
  \bibinfo {author} {\bibfnamefont {D.}~\bibnamefont {Smeets}}, \bibinfo
  {author} {\bibfnamefont {M.}~\bibnamefont {Guihard}}, \bibinfo {author}
  {\bibfnamefont {P.}~\bibnamefont {Brommer}}, \bibinfo {author} {\bibfnamefont
  {J.-F.}\ \bibnamefont {Joly}}, \bibinfo {author} {\bibfnamefont {J.-C.}\
  \bibnamefont {Pothier}}, \bibinfo {author} {\bibfnamefont {L.~J.}\
  \bibnamefont {Lewis}}, \bibinfo {author} {\bibfnamefont {N.}~\bibnamefont
  {Mousseau}}, \ and\ \bibinfo {author} {\bibfnamefont {F.}~\bibnamefont
  {Schiettekatte}},\ }\bibfield  {title} {\enquote {\bibinfo {title} {Replenish
  and relax: Explaining logarithmic annealing in ion-implanted c-{Si}},}\
  }\href@noop {} {\bibfield  {journal} {\bibinfo  {journal} {Phys. Rev. Lett.}\
  }\textbf {\bibinfo {volume} {111}} (\bibinfo {year} {2013})}\BibitemShut
  {NoStop}%
\bibitem [{\citenamefont {Jackson}(1980)}]{Jackson1980}%
  \BibitemOpen
  \bibfield  {author} {\bibinfo {author} {\bibfnamefont {J.~J.}\ \bibnamefont
  {Jackson}},\ }\bibfield  {title} {\enquote {\bibinfo {title} {Calorimeter for
  simultaneous, low‐temperature measurements of energy release and of
  resistivity recovery in irradiated metals},}\ }\href {\doibase
  10.1063/1.1136052} {\bibfield  {journal} {\bibinfo  {journal} {Rev. Sci.
  Instrum.}\ }\textbf {\bibinfo {volume} {51}},\ \bibinfo {pages} {35--41}
  (\bibinfo {year} {1980})}\BibitemShut {NoStop}%
\bibitem [{\citenamefont {Béland}\ \emph {et~al.}(2015)\citenamefont
  {Béland}, \citenamefont {Osetsky}, \citenamefont {Stoller},\ and\
  \citenamefont {Xu}}]{Beland2015}%
  \BibitemOpen
  \bibfield  {author} {\bibinfo {author} {\bibfnamefont {L.~K.}\ \bibnamefont
  {Béland}}, \bibinfo {author} {\bibfnamefont {Y.~N.}\ \bibnamefont
  {Osetsky}}, \bibinfo {author} {\bibfnamefont {R.~E.}\ \bibnamefont
  {Stoller}}, \ and\ \bibinfo {author} {\bibfnamefont {H.}~\bibnamefont {Xu}},\
  }\bibfield  {title} {\enquote {\bibinfo {title} {Slow relaxation of
  cascade-induced defects in {F}e},}\ }\href@noop {} {\bibfield  {journal}
  {\bibinfo  {journal} {Phys. Rev. B}\ }\textbf {\bibinfo {volume} {91}}
  (\bibinfo {year} {2015})}\BibitemShut {NoStop}%
\end{thebibliography}%


\end{document}